\documentclass[pdflatex,sn-mathphys-num]{sn-jnl}

\usepackage{graphicx}%
\usepackage{multirow}%
\usepackage{amsmath,amssymb,amsfonts}%
\usepackage{amsthm}%
\usepackage{mathrsfs}%
\usepackage[title]{appendix}%
\usepackage{xcolor}%
\usepackage{textcomp}%
\usepackage{manyfoot}%
\usepackage{booktabs}%
\usepackage{algorithm}%
\usepackage{algorithmicx}%
\usepackage{algpseudocode}%
\usepackage{listings}%
\usepackage{bm}
\usepackage{natbib}
\usepackage{subcaption}
\usepackage{caption}
\usepackage{anyfontsize}
\usepackage{siunitx}
\usepackage{comment}
\usepackage{tikz}
\usetikzlibrary{shapes.geometric, arrows.meta, positioning, calc}
\usetikzlibrary{fit, backgrounds}
\pgfdeclarelayer{background}
\pgfsetlayers{background,main}
\usepackage{bbm}
\usepackage{longtable}

\theoremstyle{thmstyleone}%
\newtheorem{proposition}{Proposition}[section]% 

\theoremstyle{thmstyletwo}%
\newtheorem{remark}{Remark}[section]%

\theoremstyle{thmstylethree}%

\definecolor{consolebg}{RGB}{248,248,248}
\definecolor{consoleframe}{RGB}{200,200,200}

\lstdefinestyle{matlabconsole}{
    basicstyle=\fontsize{6.5}{8}\ttfamily, % Reduces font size so lines fit naturally
    backgroundcolor=\color{consolebg},
    frame=single,
    rulecolor=\color{consoleframe},
    breaklines=true,                  % Wraps long lines
    breakatwhitespace=false,          % Wraps mid-word if needed to prevent margin overflow
    keepspaces=true,
    showstringspaces=false,
    columns=fullflexible,             % Avoids wide tracking/spacing issues
    xleftmargin=0pt,
    xrightmargin=0pt
}

\begin{document}

\title{An Integrated Framework for Assessing Interpretability in Exploratory Factor Analysis}

\author[1]{\fnm{Mariaelena} \sur{Bottazzi Schenone}}\equalcont{Authors listed in alphabetical order.}
\author*[1]{\fnm{Tiziano} \sur{Iannaccio}}\email{tiziano.iannaccio@uniroma1.it}\equalcont{Authors listed in alphabetical order.}
\author[1]{\fnm{Ilaria} \sur{Mozzetta}}\equalcont{Authors listed in alphabetical order.}
\author[1]{\fnm{Maurizio} \sur{Vichi}}\equalcont{Authors listed in alphabetical order.}

\affil*[1]{\orgdiv{Department of Statistical Sciences}, \orgname{Sapienza University of Rome}, \orgaddress{\city{Rome}, \country{Italy}}}
\abstract{Interpretability in Exploratory Factor Analysis is characterized by the fundamental trade-off between structural sparseness and explained common variance. Lacking formal evaluation metrics, researchers frequently sacrifice parsimony in the pattern matrix to maximize the model's explanatory coverage. For this reason, we extend Hoyer's sparseness ratio into a flexible \texorpdfstring{$\mathcal{L}_{p_1}/\mathcal{L}_{p_2}$}{Lp1/Lp2} formulation to introduce the Interpretability Index $I$, defined as the geometric mean of explained common variance and the generalized sparseness index. We analyze the asymptotic behavior of $I$ as norm orders increase, establishing that raising $p_2$ alone yields a stable limit, whereas increasing $p_1$ and $p_2$  jointly causes the measure to saturate. The canonical choice $(p_1,p_2)=(2,4)$ balances numerical stability with a smooth, tractable objective. We finally demonstrate, on five benchmark datasets from the statistical literature, how the Interpretability Index provides a systematic criterion for selecting the optimal parametrization of the latent space according to the user's analytical goals.}


\maketitle

\section{Introduction}
\label{sec:intro}

In Exploratory Factor Analysis (EFA) models, researchers must balance two intertwined objectives: maximizing the proportion of common variance explained by the latent factors and ensuring that said factors have a simple structure and are readily interpretable in terms of the observed variables. Pursuing the former goal often produces dense factor loading matrices with many cross-loadings, which hamper clear substantive interpretation. Conversely, forcing near-disjoint loadings can erode the overall explained common variance. 

This trade-off has shaped much of the methodological literature. Historically, interpretability in factor models has been linked to a notion of ``\textit{simple structure}'' \citep{thurstone1937psychology}, which envisions factors defined by a few salient loadings per variable alongside near-zero cross-loadings to allow for unambiguous labeling. To achieve this goal, a variety of analytic rotation techniques have been established. Orthogonal rotations constrain components to remain uncorrelated, as exemplified by \textit{Varimax} \citep{kaiser1958}, which maximizes the variance of squared loadings within columns, and the Crawford--Ferguson family \citep{CrawfordFerguson1970}, which includes methods like \textit{Quartimax} and \textit{Equamax} to trade off row- versus column-wise simplicity.

Later developments translated this conceptual target into a formal modeling framework. Confirmatory Factor Analysis (CFA) and target-rotation approaches \citep{joreskog1969general,joreskog1978structural} achieve simple structure by imposing explicit zero (or near-zero) constraints directly on the factor loadings prior to estimation. Consequently, interpretability shifted from an exploratory goal into a directly testable structural hypothesis.

However, CFA requires researchers to specify the exact loading structure \textit{a priori}. Since such knowledge is frequently unavailable, modern high-dimensional approaches have conceptually reframed the pursuit of simple structure as a problem of sparseness, where cross-loadings are penalized during the estimation process. Foundational work in this area includes penalized maximum likelihood approaches that augment the objective function with an $\mathcal{L}_1$-norm penalty, computed via a generalized Expectation-Maximization (EM) algorithm, forcing small, theoretically irrelevant cross-loadings exactly to zero during estimation \citep{choi2010penalized}.
Shortly after, \cite{ning2011sparse} incorporated $\mathcal{L}_1$-penalties directly into maximum likelihood covariance optimization without relying on EM.
However, because standard $\mathcal{L}_1$ regularization tends to over-shrink large coefficients, later extensions integrated non-concave penalties, which selectively shrink negligible loadings while leaving the magnitude of primary loadings unchanged \citep{hirose2015sparse}. 

Adapting such penalization strategies to exploratory models is mathematically complex due to the interplay between the loading matrix and the unique variances of the observed variables. To address this, \cite{trendafilov2017sparse} developed a specialized Sparse Exploratory Factor Analysis framework that utilizes matrix reparameterization to optimally locate zero-loadings without artificially inflating unique variances.

Parallel to these frequentist advancements, Bayesian frameworks enforce sparseness using spike-and-slab or global--local shrinkage priors \citep{carvalho2008high, bhattacharya2011sparse}. 

To evaluate the resulting representations, a variety of diagnostic metrics have been proposed, yet each addresses the problem only partially. General simplicity indices \citep{bentler1977factor, lorenzo2003factor} or generic sparseness metrics \citep{hoyer04, hurley09} quantify loading concentration but ignore the interplay with the covariance structure. Moreover, while regularization hyper-parameters effectively manage the trade-off between common variance extraction and structural simplicity during estimation, their values are scale-dependent and tied to specific penalty functions, meaning they cannot serve as standardized scores for comparing models across different frameworks \citep{trendafilov2014simple, hirose2015sparse,trendafilov2017sparse}.
Consequently, current proposals lack a unified, scale-free metric capable of simultaneously evaluating explained common variance and structural simplicity across different dimensions, rotations and estimation methods. To address these limitations, this paper develops a unified framework to measure interpretability in EFA.

Rather than viewing interpretability solely as a descriptive property of an estimated solution, we treat it as a measurable quantity and analyze its theoretical properties. In doing so, we propose two complementary contributions. First, we define a model-agnostic Interpretability Index, which we will denote $I$, that balances Explained Common Variance (ECV) and structural simplicity, evaluated through a novel generalized ratio, which is the second contribution of this work. Unlike existing metrics, the new index employs an imbalance-penalizing formulation, which is implemented via a geometric mean of the above metrics. A high overall score can only be achieved when a model excels in both dimensions. To quantify structural simplicity, we extend Hoyer's $\mathcal{L}_1/\mathcal{L}_2$ ratio to measure sparseness into a generalized $\mathcal{L}_{p_1}/\mathcal{L}_{p_2}$ framework that effectively suppresses negligible cross-loadings while rewarding concentrated loading patterns. To complement this global model-level evaluation with item-level diagnostics, we also establish a set of complementary, row-wise structural compliance indices inspired by classical simple structure principles \citep{thurstone1947}. 

In this paper: Section \ref{sec:measuring_interpretability} defines the Interpretability Index $I$ and all its constituents. Section \ref{sec:asymptotics} analyses the asymptotic behavior of the proposed index and the rationale for the canonical choice of $p_1$ and $p_2$. Section~\ref{sec:properties} presents the theoretical properties of $I$, while Section~\ref{sec:workflow_software} provides a practical workflow for its use, together with an accompanying MATLAB routine that automates the full pipeline. Section \ref{sec:thurstone} introduces a set of complementary, row-wise structural compliance indices to use alongside $I$ as a robustness diagnostic across candidate solutions. Section \ref{sec:applications} applies the framework to real-world data, and Section \ref{sec:conclusions} provides conclusions.

% =========================================================================
\section{Measuring Interpretability}
\label{sec:measuring_interpretability}
% =========================================================================

Consider the Exploratory Factor Analysis (EFA) model:
\begin{equation}
    \label{eq1}
    \mathbf{X} = \mathbf{F\Lambda^\top}+ \mathbf{E},
\end{equation}
where $\mathbf{X}$ is the $(N \times J)$ units by variables data matrix, $\mathbf{F}$ is the $(N \times Q)$ factor matrix ($Q \leq J$), $\mathbf{\Lambda}$ is the $(J \times Q)$ pattern/loading matrix, and $\mathbf{E}$ is the $(N \times J)$ error matrix. The $(J \times J)$ sample covariance matrix $\mathbf{S}$ is modeled as follows:
\begin{equation}
    \label{eq2}
    \hat{\mathbf{\Sigma}} = \mathbf{\Lambda \Phi \Lambda^\top}+\mathbf{\Psi},
\end{equation}
where $\mathbf{\Phi}\succeq 0$ is the $(Q \times Q)$ Positive Semi-Definite (PSD) factor covariance matrix and $\mathbf{\Psi}\succ 0$ is the PD diagonal $(J \times J)$ uniqueness matrix, both symmetric.

Interpretability depends on both the fraction of common factor variance captured and how cleanly the loadings are distributed. For this reason, given two norm orders $1 \leq p_1 < p_2 < \infty$, the Interpretability Index $I(\mathbf{\Lambda}, \mathbf{\Phi}, \mathbf{D}, p_1, p_2)$ is quantified as follows:
\begin{equation}
    \label{eq3}
    I(\mathbf{\Lambda}, \mathbf{\Phi},\mathbf{D}, p_1, p_2) = \sqrt{\; \operatorname{ECV}(\mathbf{\Lambda},\mathbf{\Phi},\mathbf{D}) \cdot \overline{\mathrm{GS}}_w(\mathbf{\Lambda},\mathbf{\Phi}, p_1, p_2)\;} \in [0,1].
\end{equation}

In this formulation, the proportion of common variance accounted for is:
\begin{equation}
    \label{eq4b}
    \operatorname{ECV}(\mathbf{\Lambda},\mathbf{\Phi},\mathbf{D}) = \frac{\operatorname{tr}(\mathbf{D}^{-1/2}(\mathbf{\Lambda\Phi\Lambda^\top})\mathbf{D}^{-1/2})}{J} \in [0,1],
\end{equation}
where $\mathbf{D} = \operatorname{diag}(\hat{\mathbf{\Sigma}})$. 

If the model is fit on the sample correlation matrix $\mathbf{R}$ and thus reproduces the diagonal $(\hat{\boldsymbol{\Sigma}}_{jj}=1,\ j=1,\ldots,J)$, then $\mathbf{D}=\mathbf{I}_J$ and Equation \eqref{eq4b} reduces to $\frac{1}{J}\operatorname{tr}(\mathbf{\Lambda \Phi \Lambda^\top})$.

On the other hand, the ``simplicity'' of the model's solution is quantified through:
\begin{equation}
    \label{eq5}   
    \overline{\mathrm{GS}}_w(\mathbf{\Lambda},\mathbf{\Phi}, p_1, p_2) = \mathbf{w}(\mathbf{\Lambda},\mathbf{\Phi})^\top \mathbf{gs}(\mathbf{\Lambda},p_1,p_2) \in [0,1],
\end{equation}
i.e., the weighted average formed by the weight vector:
\begin{equation}
    \label{eq:w_orth_vector}
    \mathbf{w}(\mathbf{\Lambda},\mathbf{\Phi}) = \frac{\operatorname{diag}\left( \mathbf{\Phi}^{1/2} \mathbf{\Lambda}^\top \mathbf{\Lambda} \mathbf{\Phi}^{1/2} \right)}{\operatorname{tr}\left( \mathbf{\Lambda} \mathbf{\Phi}\mathbf{\Lambda}^\top \right)}
\end{equation}
and the sparseness vector $\mathbf{gs}(\mathbf{\Lambda},p_1,p_2)$ with individual factor entries:
\begin{equation}
    \label{eq6}
    \mathrm{gs}(\boldsymbol{\lambda}_q, p_1, p_2)= \frac{J^{\left(\frac{1}{p_1} - \frac{1}{p_2}\right)} - \frac{\|\boldsymbol{\lambda}_q\|_{p_1}}{\|\boldsymbol{\lambda}_q\|_{p_2}}}{J^{\left(\frac{1}{p_1} - \frac{1}{p_2}\right)} - 1},    
\end{equation}
where $\|\boldsymbol{\lambda}_q\|_{p} = \left( \sum_{j=1}^J |\lambda_{jq}|^p \right)^{1/p}$.

To clarify each constituent of $\overline{\mathrm{GS}}_w$, we devote Subsection \ref{sec:weighting_scheme} to the choice of an appropriate weight system, Subsection \ref{sec:generalized_sparseness} to the quantification of ``simplicity'' through sparseness, and Section \ref{sec:asymptotics} to the canonical choice of $p_1$ and $p_2$.

% -------------------------------------------------------------------------
\subsection{Weighting scheme for aggregating sparseness}
\label{sec:weighting_scheme}
% -------------------------------------------------------------------------

To avoid inflating $I$ when many weak simple factors coexist with few dominant dense factors, $\overline{\mathrm{GS}}_w$ weights each factor's share of common variance through vector $\mathbf{w}$.

\paragraph{General case: oblique factors}
Under oblique structures, the allocation of common variance across factors is non-trivial due to inter-factor correlations. To isolate each factor's unique contribution, we adopt the orthogonalized weight vector $\mathbf{w}(\mathbf{\Lambda}, \mathbf{\Phi})$ introduced in Equation \eqref{eq:w_orth_vector}.

Geometrically, the principal inverse square root $\mathbf{\Phi}^{-1/2}$ performs symmetric (Mahalanobis) whitening of the factor space. Specifically, defining $\mathbf{\Gamma} = \mathbf{\Lambda\Phi}^{1/2}$ gives $\mathbf{\Lambda}\mathbf{\Phi}\mathbf{\Lambda}^\top = \mathbf{\Gamma\Gamma}^\top$. Thus, $\mathbf{\Gamma}$ provides an equivalent orthogonal representation of the model-implied common covariance structure (here we assume $\mathbf{\Phi}\succ 0$).

Notably, this formulation yields an \emph{exact partition} of the explained common variance. By the cyclic property of the trace, the sum of the unnormalized weights matches the numerator of $\operatorname{ECV}$ (after standardization to the correlation scale). 

Furthermore, the resulting weights are non-negative, sum to 1, and are invariant to permutations and sign flips of the columns of $\mathbf{\Lambda}$.

\paragraph{Special case: orthogonal factors}
When $\mathbf{\Phi} = \mathbf{I}_Q$, factor whitening has no effect and Equation \eqref{eq:w_orth_vector} reduces to:
\begin{equation}
    \label{eq:w_dir_vec}
    \mathbf{w}^{(\mathrm{dir})}(\mathbf{\Lambda}) = \mathbf{w}(\mathbf{\Lambda}, \mathbf{I}_Q) = \frac{\operatorname{diag}\left( \mathbf{\Lambda}^\top \mathbf{\Lambda} \right)}{\operatorname{tr}\left( \mathbf{\Lambda}^\top \mathbf{\Lambda} \right)}.
\end{equation}
Here, the $q$-th element $w_q^{(\mathrm{dir})} = \frac{\|\boldsymbol{\lambda}_q\|_2^2}{\sum_{r=1}^Q \|\boldsymbol{\lambda}_r\|_2^2}$ represents the proportion of common variance directly attributable to factor $q$'s unadjusted loadings. 

\begin{remark}[Connecting EFA to PPCA]
Probabilistic PCA (PPCA) is obtained as a special case of the general EFA model \eqref{eq2} by enforcing $\mathbf{\Phi}=\mathbf{I}_Q$ and $\mathbf{\Psi}=\sigma^2\mathbf{I}_J$:
\begin{equation}
    \hat{\boldsymbol{\Sigma}}=\mathbf{\Lambda\Lambda}^\top+\sigma^2\mathbf{I}_J.
\end{equation}
On the correlation scale ($\mathbf{D}=\mathbf{I}_J$), PPCA yields:
\begin{equation}
    \operatorname{ECV}=\frac{1}{J}\operatorname{tr}(\mathbf{\Lambda\Lambda}^\top) = \frac{1}{J}\sum_{q=1}^Q \|\boldsymbol{\lambda}_q\|_2^2,
\end{equation}
which represents the average communality captured by the $Q$ latent dimensions. Since PPCA assumes isotropic error variance across all variables ($\psi_j = \sigma^2$ for all $j$), it restricts the model's ability to accommodate variable-specific noise levels. 

In contrast, general EFA relaxes both constraints simultaneously. First, heteroskedastic uniqueness ($\mathbf{\Psi} = \operatorname{diag}(\psi_1,\dots,\psi_J)$) allows each observed variable to possess a distinct unique variance, ensuring that $\operatorname{ECV}<1$ whenever $\operatorname{tr}(\mathbf{\Psi}) > 0$. Second, oblique factor structures allow correlated latent attributes, leaving the resulting common covariance $\mathbf{\Lambda}\mathbf{\Phi}\mathbf{\Lambda}^\top$ invariant under non-singular factor transformations. Thus, PPCA serves as the orthogonal, isotropic baseline within the broader EFA family.
\end{remark}

% -------------------------------------------------------------------------
\subsection{A generalized measure of sparseness}
\label{sec:generalized_sparseness}
% -------------------------------------------------------------------------

To quantify the sparseness of a vector $\mathbf{v} = (v_1, \dots, v_d)^\top \in \mathbb{R}^d$, \cite{hoyer04} leveraged the fact that the $\mathcal{L}_2$ norm decreases more rapidly than the $\mathcal{L}_1$ norm as the elements of $\mathbf{v}$ become more concentrated. By scaling the ratio of these norms, Hoyer was able to map $\mathbf{v}$ onto the unit interval. This measure assigns $0$ to uniform vectors and $1$ to vectors containing only a single non-zero element:
\begin{equation}
    \mathrm{sparseness}(\mathbf{v}) = \frac{\sqrt{d} - \frac{\sum_{i=1}^d |v_i|}{\sqrt{\sum_{i=1}^d v_i^2}}}{\sqrt{d} - 1}.
    \label{eq:hoyer}
\end{equation}

While Hoyer's measure captures concentration in a purely mathematical way, it was not specifically designed with latent variable modeling in mind. A loading vector is indeed perceived as highly interpretable if it is dominated by a single primary variable. However, the $\mathcal{L}_1/\mathcal{L}_2$ ratio is overly sensitive to ``noise'' (in our case, cross-loadings). The presence of several negligible but non-zero values can significantly deflate sparseness. For instance, a vector in which $95\%$ of the mass is concentrated in a single entry, i.e., an ideal structure for most statisticians, scores at around $0.7$, failing to register the high clarity of the loading pattern.

To adapt Hoyer's measure to our goals, we propose the generalized $\mathcal{L}_{p_1}/\mathcal{L}_{p_2}$ sparseness ($\mathrm{gs}(\cdot,p_1,p_2)$) shown in \eqref{eq6}. Hoyer’s measure is recovered by setting $p_1=1$ and $p_2=2$:
\begin{equation}
    \label{eq:gs}
    \mathrm{gs}(\mathbf{v}, 1, 2) = \frac{d^{1 - \frac{1}{2}} - \frac{\|\mathbf{v}\|_1}{\|\mathbf{v}\|_2}}{d^{1 - \frac{1}{2}} - 1} = \frac{\sqrt{d} - \frac{\sum_{i=1}^d |v_i|}{\sqrt{\sum_{i=1}^d v_i^2}}}{\sqrt{d} - 1}.
\end{equation}

This formulation offers great flexibility in penalizing negligible elements while preserving the desirable features of the original measure. In particular:
\begin{equation}
    \label{eq:bound}
    \mathrm{gs}(\cdot,p_1,p_2) \in [0,1] \quad \forall \; 1 \leq p_1 < p_2 < \infty.
\end{equation}

The result above is quickly proved by showing that:
\begin{equation}
    \label{eq:normbounds}
    d^{\left(\frac{1}{p_1} - \frac{1}{p_2}\right)}\|\mathbf{v}\|_{p_2} \;\; \geq \;\; \|\mathbf{v}\|_{p_1} \;\; \geq \;\; \|\mathbf{v}\|_{p_2}.
\end{equation}

The lower bound of the above inequality is a consequence of the monotonicity of $\mathcal{L}_p$ norms with respect to $p$. The upper bound is derived from H\"older's inequality, where the constant $d^{(1/p_1 - 1/p_2)}$ serves as the necessary normalization factor. Furthermore, since $d^{(1/p_1 - 1/p_2)}$ is an increasing function of $(1/p_1 - 1/p_2)$, higher-order ratios downplay the contribution of smaller loadings, thus better mirroring what analysts perceive as ``simple''. These bounds ensure that $\mathrm{gs}(\cdot,p_1,p_2)=1$ for a vector with a single non-zero entry (maximum sparseness), and $\mathrm{gs}(\cdot,p_1,p_2)=0$ for a uniform vector where all entries are equal (minimum sparseness).

% =========================================================================
\section{Asymptotic behavior of \texorpdfstring{$p_1,p_2$}{p1,p2}}
\label{sec:asymptotics}
% =========================================================================

To guide parameter choice and establish theoretical boundaries, we analyze the behavior of $\mathrm{gs}(\cdot, p_1, p_2)$ under two different limiting regimes as $p_1, p_2 \to \infty$. 

Consider a vector $\mathbf{v} \in \mathbb{R}^d$ with a unique largest-magnitude entry. We arrange the absolute values of its components in descending order:
\begin{equation}
    |v_{(1)}| > |v_{(2)}| \ge \dots \ge |v_{(d)}| \ge 0.
\end{equation}
Let $v_{\max} = \vert{}v_{(1)}\vert{}$ denote the maximum magnitude, and define the relative runner-up ratios as:
\begin{equation}
    \label{runnerup}
    r_k = \frac{|v_{(k)}|}{v_{\max}} \in [0, 1), \quad \text{for } k = 2, \dots, d.
\end{equation}
Using these ratios, the $\mathcal{L}_p$-norm of $\mathbf{v}$ can be factorized as:
\begin{equation}
    \label{normexp}
    \|\mathbf{v}\|_p = v_{\max}\left(1 + T_p\right)^{1/p}, \quad \text{where } T_p := \sum_{k=2}^d r_k^{\,p}.
\end{equation}
Because each $r_k < 1$, the term $T_p$ decays exponentially fast in $p$ at a rate governed primarily by $r_2$ (the slowest-decaying term). Consequently:
\begin{equation}
    \lim_{p \to \infty} \|\mathbf{v}\|_p = v_{\max} = \|\mathbf{v}\|_\infty. 
\end{equation} 

The result above is the core mechanism behind both the convergence results in the next two subsections. 

\begin{remark} 
If the maximum magnitude is attained with multiplicity $m > 1$, the leading factor $v_{\max}$ is replaced by $m^{1/p}v_{\max}$. Since $m^{1/p} \to 1$ as $p \to \infty$, the asymptotic conclusions remain unchanged.
\end{remark}

% -------------------------------------------------------------------------
\subsection{Behavior under fixed \texorpdfstring{$p_1$}{p1} and growing \texorpdfstring{$p_2$}{p2} (\texorpdfstring{$\mathcal{L}_{p_1}/\mathcal{L}_\infty$}{Lp1/Linfty} Limit)}
\label{sec:limit_p2}
% -------------------------------------------------------------------------

\begin{figure}[h!]
    \centering
    \includegraphics[width=0.7\textwidth]{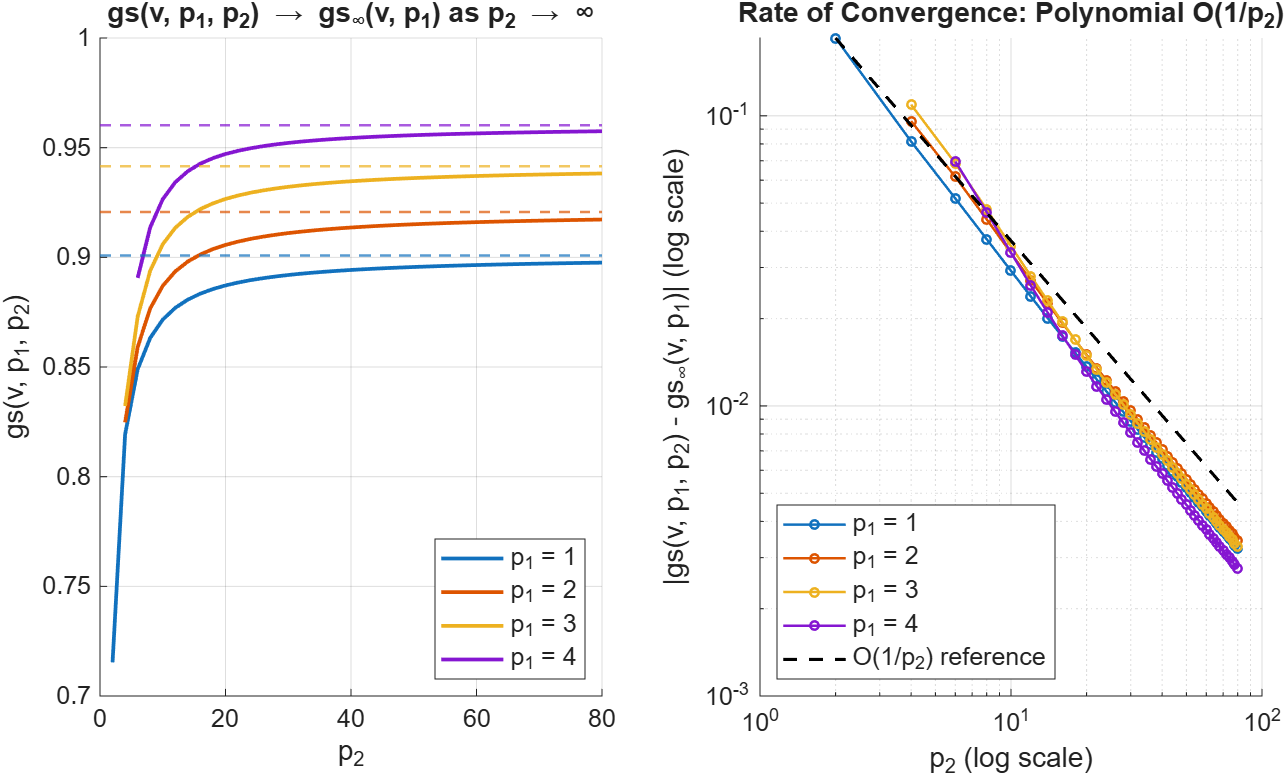}
    \caption{Validation of Proposition \ref{prop:limit_p2} on a sample loading vector ($d=10$, $v_{\max}=1$, runner-up ratio $r_2=0.6$). Left: $\mathrm{gs}(\mathbf{v}, p_1, p_2)$ as a function of $p_2$ for fixed $p_1 \in \{1, 2, 3, 4\}$ (solid lines), converging smoothly to their respective $\mathcal{L}_{p_1}/\mathcal{L}_\infty$ limits $\mathrm{gs}_\infty(\mathbf{v}, p_1)$ (dashed lines). Right: Log-log plot of the approximation error, confirming the theoretical $\mathcal{O}(1/p_2)$ polynomial decay against the reference slope (black dashed line).}
    \label{fig:1}
\end{figure}

Keeping $p_1 \ge 1$ fixed while $p_2 \to \infty$ yields a well-defined $\mathcal{L}_{p_1}/\mathcal{L}_\infty$ limiting index with polynomial error, as visualized in Figure \ref{fig:1}.

\begin{proposition}[$\mathcal{L}_{p_1}/\mathcal{L}_\infty$ Limit and Convergence Rate]
\label{prop:limit_p2}
For any fixed $p_1 \ge 1$, as $p_2 \to \infty$,
\begin{equation}
    \lim_{p_2 \to \infty} \mathrm{gs}(\mathbf{v}, p_1, p_2) \;=\; \mathrm{gs}_\infty(\mathbf{v}, p_1) \;:=\; \frac{d^{1/p_1} - \|\mathbf{v}\|_{p_1}/v_{\max}}{d^{1/p_1} - 1},
    \label{eq:caseA}
\end{equation}
where $\mathrm{gs}_\infty(\mathbf{v}, p_1) \in [0, 1]$ is itself a valid sparseness index. Furthermore, the convergence toward this limit proceeds at a polynomial rate $\mathcal{O}(1/p_2)$ (see Figure \ref{fig:1}):
\begin{equation}
    \mathrm{gs}(\mathbf{v}, p_1, p_2) \;=\; \mathrm{gs}_\infty(\mathbf{v}, p_1) \;+\; \frac{d^{1/p_1}\ln d}{p_2\bigl(d^{1/p_1}-1\bigr)}\Bigl(\mathrm{gs}_\infty(\mathbf{v}, p_1) - 1\Bigr) \;+\; \mathcal{O}\!\left(\frac{1}{p_2^2}\right).
    \label{eq:caseA-rate}
\end{equation}
\end{proposition}

\begin{remark} 
If $m > 1$, the convergence rate remains $\mathcal{O}(1/p_2)$, with \eqref{eq:caseA-rate} taking the generalized form:
\begin{equation}
    \mathrm{gs}(\mathbf{v}, p_1, p_2) = \mathrm{gs}_\infty(\mathbf{v}, p_1) + \frac{d^{1/p_1}\ln d \bigl(\mathrm{gs}_\infty(\mathbf{v}, p_1) - 1\bigr) + \frac{\|\mathbf{v}\|_{p_1}\ln m}{v_{\max}}}{p_2\bigl(d^{1/p_1}-1\bigr)} + \mathcal{O}\!\left(\frac{1}{p_2^2}\right).
    \label{eq:caseA-rate-multiplicity}
\end{equation}
\end{remark}

% -------------------------------------------------------------------------
\subsection{Behavior under jointly growing \texorpdfstring{$p_1$}{p1} and \texorpdfstring{$p_2$}{p2}}
\label{sec:joint_growth}
% -------------------------------------------------------------------------

\begin{figure}[ht!]
    \centering
    \includegraphics[width=0.7\textwidth]{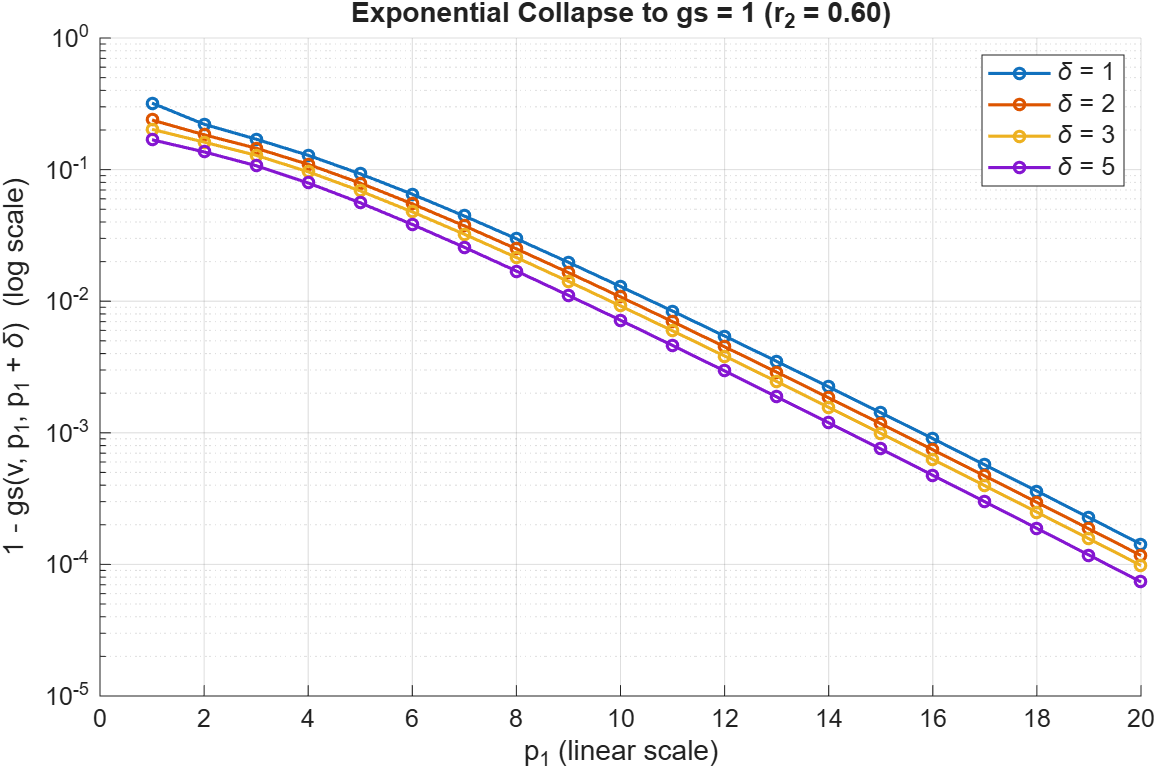}
    \caption{Validation of Proposition \ref{prop:saturation} on a sample loading vector ($d=10$, $r_2=0.6$). $1 - \mathrm{gs}(\mathbf{v}, p_1, p_1+\delta)$ vs.\ $p_1$ on a semi-log scale for fixed gaps $\delta \in \{1, 2, 3, 5\}$. The linear decay on the logarithmic scale confirms the rapid exponential collapse toward $\mathrm{gs} = 1$ predicted by \eqref{eq:caseB-rate}, irrespective of the gap $\delta$.}
    \label{fig:2}
\end{figure}

Next, we examine the scenario where $p_1$ and $p_2$ are increased simultaneously while maintaining a fixed gap $\delta = p_2 - p_1 > 0$. The measure will collapse to unity regardless of the vector's underlying sparseness profile, as visualized in Figure \ref{fig:2}.

\begin{proposition}[Joint Growth and Exponential Saturation]
\label{prop:saturation}
Fix a gap $\delta = p_2 - p_1 > 0$ and let $p_1 \to \infty$. Then, for any vector $\mathbf{v} \in \mathbb{R}^d$ possessing a unique dominant entry,
\begin{equation}
    \lim_{p_1 \to \infty} \mathrm{gs}(\mathbf{v}, p_1, p_1 + \delta) \;=\; 1.
    \label{eq:caseB}
\end{equation}
Furthermore, $1 - \mathrm{gs}(\mathbf{v}, p_1, p_1 + \delta)$ decays exponentially fast:
\begin{equation}
    1 - \mathrm{gs}(\mathbf{v}, p_1, p_1 + \delta) \;=\; \mathcal{O}\!\left(\frac{p_1^2}{\delta \ln d}\; r_2^{\,p_1}\right) \;\xrightarrow[p_1 \to \infty]{}\; 0.
    \label{eq:caseB-rate}
\end{equation}
\end{proposition}

\begin{remark}
\label{rem:multiplicity_saturation}
If $m > 1$, \eqref{eq:caseB} generalizes to:
\begin{equation}
    \lim_{p_1 \to \infty} \mathrm{gs}(\mathbf{v}, p_1, p_1 + \delta) = 1 - \frac{\ln m}{\ln d},
    \label{eq:limit_multiplicity}
\end{equation}
which corresponds to the generalized sparseness of a flat $m$-sparse vector in $\mathbb{R}^d$.
\end{remark}

The asymptotic behavior above is characterized by two distinct mechanisms: (1) the influence of all non-maximal components $|v_{(k)}| < v_{\max}$ vanishes exponentially fast at rate $\mathcal{O}\left(\frac{p_1^2}{\delta \ln d} \, r_{m+1}^{p_1}\right)$; (2) once the tail components vanish, $\mathrm{gs}$ approaches the limit $\frac{\ln(d/m)}{\ln d}$ at rate $\mathcal{O}(1/p_1)$.

\begin{proposition}[Rank Stability along $p_2 \to \infty$]
\label{prop_rank}
Fix $p_1 \ge 1$. For any loading vectors $\mathbf{v}, \mathbf{v}'$ satisfying $\mathrm{gs}_\infty(\mathbf{v}, p_1) > \mathrm{gs}_\infty(\mathbf{v}', p_1)$, there exists a finite threshold $p_2^*$ such that for all $p_2 \ge p_2^*$,
\begin{equation*}
    \mathrm{gs}(\mathbf{v}, p_1, p_2) > \mathrm{gs}(\mathbf{v}', p_1, p_2).
\end{equation*}
\end{proposition}

\begin{remark}[Pathway Choice]
Proposition \ref{prop_rank} provides a practical guarantee: when evaluating sparseness along the $p_2 \to \infty$ trajectory with fixed $p_1$, finite choices of $p_2$ preserve the true asymptotic ordering of candidate factor models. This stability breaks down along diagonal trajectories ($p_1 \to \infty, p_2 = p_1 + \delta$), where all models with a unique dominant loading collapse to the same limiting value of 1.
\end{remark}

% -------------------------------------------------------------------------
\subsection{Implications for Parameter Boundaries}
\label{sec:implications_boundaries}
% -------------------------------------------------------------------------

\begin{figure}[ht!]
    \centering
    \includegraphics[width=0.8\textwidth]{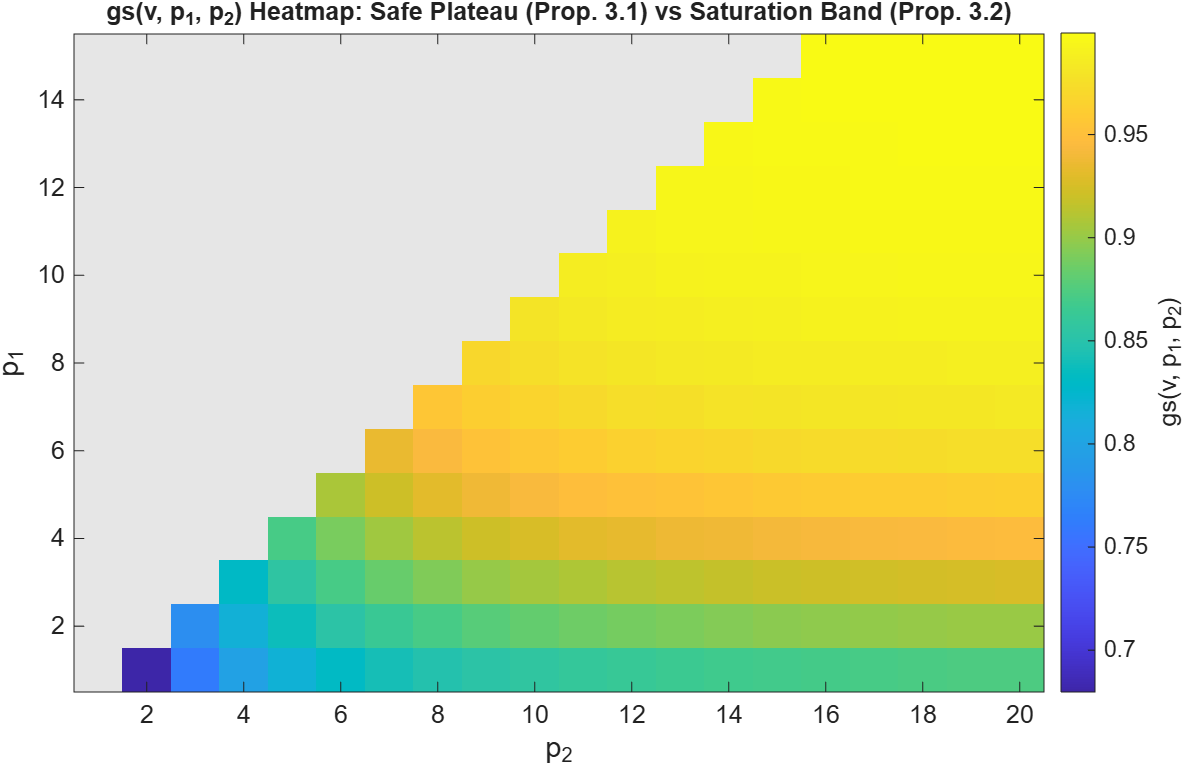}
    \caption{Heatmap of $\mathrm{gs}(\mathbf{v}, p_1, p_2)$ across the parameter domain $1 \le p_1 < p_2 \le 20$. Moving horizontally (increasing $p_2$ at fixed $p_1$) leads to smooth $\mathcal{O}(1/p_2)$ stabilization toward the $\mathcal{L}_{p_1}/\mathcal{L}_\infty$ ceiling (Proposition \ref{prop:limit_p2}). Moving diagonally (increasing $p_1$ and $p_2$ jointly) drives the metric into the exponential saturation region $\mathrm{gs} \approx 1$ (Proposition \ref{prop:saturation}).}
    \label{fig:gs_heatmap}
\end{figure}

The findings in Section \ref{sec:limit_p2} and Section \ref{sec:joint_growth} guide the configuration of the generalized sparseness metric. Proposition \ref{prop:limit_p2} shows that holding $p_1$ low while increasing $p_2$ provides a safe parameter trajectory; the measure converges at rate $\mathcal{O}(1/p_2)$ toward its $\mathcal{L}_{p_1}/\mathcal{L}_\infty$ limit while fully retaining its capacity to discriminate between varying sparsity profiles. 

On the other hand, Proposition \ref{prop:saturation} warns that joint growth of both parameters past a low threshold (typically $p_1 \ge 15$--$20$) pushes $\mathrm{gs}$ into saturation ($\mathrm{gs} \approx 1$) at exponential speed, regardless of the vector's true mass distribution. This makes $\mathrm{gs}$ lose its discriminatory power.

Figure \ref{fig:gs_heatmap} illustrates this dual behavior across the parameter space, contrasting the well-behaved horizontal paths against the saturation regions created by diagonal trajectories. Consequently, $p_1$ and $p_2$ cannot be tuned upward arbitrarily; $p_1$ must remain anchored at a low canonical value to ensure reliable cross-model benchmarking.

% -------------------------------------------------------------------------
\subsection{Canonical Parameter Selection: \texorpdfstring{$p_1 = 2$ and $p_2 = 4$}{p1 = 2 and p2 = 4}}
\label{sec:canonical_selection}
% -------------------------------------------------------------------------

While \eqref{eq6} holds for any valid norm pair $1 \le p_1 < p_2 < \infty$, setting $(p_1,p_2) = (2,4)$ provides key theoretical and computational advantages that make it the canonical choice for the interpretability index.

When evaluating a $d$-dimensional factor loading vector $\mathbf{v} \in \mathbb{R}^d$ under the unit-length constraint ($\|\mathbf{v}\|_2 = 1$), its squared entries $v_i^2 \in [0, 1]$ represent a discrete probability-like allocation of the factor's SS loadings. 

On one hand, (1) the least interpretable baseline occurs when factor variance is evenly distributed across all coordinates ($v_i = 1/\sqrt{d}$ for all $i = 1, \dots, d$), scoring minimal kurtosis (platykurtic behavior). On the other hand, (2) the ideal structure occurs when the factor concentrates its entire variance onto a single coordinate ($v_i = 1$ and $v_{j \neq i} = 0$), corresponding to a degenerate distribution with maximal kurtosis (leptokurtic behavior).

Setting $p_1 = 2$ significantly simplifies the analytical expression of $\mathrm{gs}(\mathbf{v}, p_1, p_2)$ by reducing the $\mathcal{L}_{p_1}$ norm to unity in case of orthonormality (the baseline scenario in EFA, with $\mathbf{\Phi}=\mathbf{I}_Q$). 

Setting $p_2=4$ links the measure to the fourth moment of the loading distribution. Under the unit-$\mathcal{L}_2$ constraint, greater concentration produces a larger $\sum_i v_i^4$, and hence a larger $\mathcal{L}_4$ norm. Consequently, $\mathrm{gs}(\mathbf{v},2,4)$ increases toward 1 as the loading mass becomes concentrated on fewer coordinates. Conversely, a vector with substantial cross-loadings exhibits low kurtosis, driving $\mathrm{gs}$ down toward 0. Under $(p_1,p_2)=(2,4)$ and $\mathbf{\Phi}=\mathbf{I}_Q$, Equation \eqref{eq6} simplifies to:
\begin{equation}
    \label{eq:canon}
    \mathrm{gs}(\boldsymbol{\lambda}_q,2,4) = \mathrm{gs}(\boldsymbol{\lambda}_q) = \frac{J^{\frac{1}{4}}-\|\boldsymbol{\lambda}_q\|_4^{-1}}{J^{\frac{1}{4}}-1}.
\end{equation}

\begin{remark}
The choice $(p_2=4)$ is further anchored in classical literature through its direct link to the Quartimax criterion \citep{NeuhausWrigley1954}. This rotation simplifies factors by maximizing the total sum of fourth-power loadings $\sum_{i} \sum_{j} v_{ij}^4 = \sum_{j} \|\mathbf{v}_j\|_4^4$. Under unit $\mathcal{L}_2$ constraints, $\mathrm{gs}(\mathbf{v}_j, 2, 4)$ is a strictly monotonic function of $\|\mathbf{v}_j\|_4^4$ for any single factor column $\mathbf{v}_j$, reducing $p_2=4$ to a measure of the exact loading concentration targeted by the fourth-order rotation.

However, performing a Quartimax rotation under $\mathbf{\Phi}=\mathbf{I}_Q$ does not automatically maximize $\overline{\mathrm{GS}}_w$. Since $\mathrm{gs}$ weights each factor by $w_q$, alternative rotations such as Varimax \citep{kaiser1958} may achieve a superior tradeoff between factor sparseness and explained common variance. Consequently, $\overline{\mathrm{GS}}_w$ can be considered a diagnostic metric to evaluate and compare candidate factor rotations objectively.
\end{remark}

% =========================================================================
\section{Properties of the Interpretability Index}
\label{sec:properties}
% =========================================================================

Under the framework introduced in Section \ref{sec:measuring_interpretability}, the Interpretability Index satisfies the following results:

\begin{proposition}[Bounds and Extrema]
\label{prop:extrema}
$I \in [0,1]$. Moreover,
\begin{align*}
    I=0 &\iff \operatorname{ECV}=0 \ \text{or}\ \overline{\mathrm{GS}}_w=0,\\
    I=1 &\iff \operatorname{ECV}=1\ \text{and}\ \overline{\mathrm{GS}}_w=1.    
\end{align*}
\end{proposition}

\begin{proof}
Since $\operatorname{ECV} \in [0,1]$ by definition and $\mathrm{gs} \in [0,1]$ as shown in Section \ref{sec:generalized_sparseness}, it follows that $\overline{\mathrm{GS}}_w \in [0,1]$ as $w_q \ge 0$ with $\sum_q w_q = 1$. Consequently, $I = \sqrt{\operatorname{ECV} \cdot \overline{\mathrm{GS}}_w} \in [0,1]$. The extrema conditions $I=0$ and $I=1$ follow immediately from the non-negativity and upper bounds of its factors.
\end{proof}

\begin{proposition}[Arithmetic Mean--Geometric Mean Non-Compensation]
\label{prop:am_gm}
\begin{equation*}
    I \le \frac{\operatorname{ECV}+\overline{\mathrm{GS}}_w}{2},
\end{equation*}
with equality holding iff $\operatorname{ECV}=\overline{\mathrm{GS}}_w$.
\end{proposition}

\begin{proof}
The result follows directly from the Arithmetic Mean--Geometric Mean (AM--GM) inequality applied to the non-negative terms $\operatorname{ECV}$ and $\overline{\mathrm{GS}}_w$.
\end{proof}

\begin{proposition}[Permutation/Sign Invariance]
\label{prop:permutation}
$I$ is invariant under (i) any permutation and (ii) any column-wise sign reflection of $\mathbf{\Lambda}$.
\end{proposition}

\begin{proof}
$\operatorname{ECV}$ is unaffected by orthogonal $\mathbf{T}$; $\overline{\mathrm{GS}}_w$ is unchanged because norms and the quadratic forms used by the weights are sign-invariant, and column permutations just reorder summands.
\end{proof}

\begin{proposition}[Monotonicity in Each Constituent]
\label{prop:monotonicity}
For fixed $\overline{\mathrm{GS}}_w>0$, the index $I$ increases strictly in $\operatorname{ECV}$; for fixed $\operatorname{ECV}>0$, $I$ increases strictly in $\overline{\mathrm{GS}}_w$.   
\end{proposition}

\begin{proof}
Since $f(x) = \sqrt{x}$ is strictly increasing on $(0, \infty)$, scaling $x$ by a positive constant preserves strict monotonicity. The result follows directly for both $\operatorname{ECV}$ and $\overline{\mathrm{GS}}_w$.
\end{proof}

\begin{proposition}[Marginal Sensitivity]
\label{prop:marginal_sensitivity}
For $\operatorname{ECV} > 0$ and $\overline{\mathrm{GS}}_w > 0$, the marginal gain of $I$ with respect to either factor is strictly greater for the smaller factor.
\end{proposition}

\begin{proof}
Direct differentiation of $I = (\operatorname{ECV} \cdot \overline{\mathrm{GS}}_w)^{1/2}$ with respect to $\operatorname{ECV}$ yields:
\begin{equation*}
    \frac{\partial I}{\partial \operatorname{ECV}} = \frac{1}{2}\sqrt{\frac{\overline{\mathrm{GS}}_w}{\operatorname{ECV}}},
\end{equation*}
with $\frac{\partial I}{\partial \overline{\mathrm{GS}}_w}$ following by symmetry. It follows directly that $\frac{\partial I}{\partial \operatorname{ECV}} > \frac{\partial I}{\partial \overline{\mathrm{GS}}_w}$ iff $\operatorname{ECV} < \overline{\mathrm{GS}}_w$.
\end{proof}

% =========================================================================
\section{Using the Interpretability Index: A Practical Workflow and Software Implementation}
\label{sec:workflow_software}
% =========================================================================

This section offers a compact workflow showing how to adopt the Interpretability Index to enhance EFA. We provide the complete MATLAB routine that implements every stage below end-to-end and returns a single, standardized report. The routine is built to be user-friendly: sensible defaults reproduce the recommended workflow out of the box, while user-specific modeling needs can be accommodated through name--value argument tuning. The routine is available at \texttt{https://github.com/IannaccioStat/Interpretability}, ensuring exact reproducibility of the results in Section \ref{sec:applications}.
Note that all numerical results reported in this paper were obtained using MATLAB R2026a, together with \texttt{R} 4.6.1 (accessed through \texttt{RStudio} 2026.8.0.187).

\paragraph{Interpretability Evaluation} 
The evaluation of model interpretability is conducted via the function \texttt{interp}, which takes as inputs the factor pattern matrix $\mathbf{\Lambda}$, the factor correlation matrix $\mathbf{\Phi}$ and a set of user-defined execution parameters. 

The function first performs a PSD check on $\mathbf{\Phi}$. If numerical instabilities are detected, $\mathbf{\Phi}$ is projected to the nearest valid PSD correlation matrix using spectral decomposition (returning a console alert). $\operatorname{ECV}$ is then evaluated on the correlation scale, alongside $\mathbf{gs}(\mathbf{\Lambda}, 2, 4)$, after which $I$ is computed according to Equation \eqref{eq3}. 

In addition, the function returns a structured MATLAB output object containing factor-specific diagnostics and an orthogonality check on $\mathbf{\Phi}$ to account for inter-factor redundancy. To do so, the orthogonality index (one minus the RMS of off-diagonal correlations) is measured as:
\begin{equation}
    \mathrm{Orth}(\mathbf{\Phi}) = 1 - \sqrt{\frac{2}{Q(Q-1)} \sum_{1 \le q < r \le Q} \phi_{qr}^2}, \qquad Q \ge 2.
    \label{eq:orth_clean}
\end{equation} 

\paragraph{Comprehensive Model Evaluation}
An automated framework designed to streamline model evaluation and rotation selection in EFA is provided here. To execute the complete pipeline, the user only needs to supply the raw data matrix $\mathbf{X}$ (containing continuous, ordinal, or mixed variables) and the desired number of factors $Q$. This input is processed by the primary driver routine \texttt{interpEval}, which automatically computes and compiles the performance metrics for both the unrotated baseline model and ten candidate factor rotations. Naturally, arbitrary rotation algorithms or alternative extraction methods can be evaluated directly via the \texttt{interp} function.

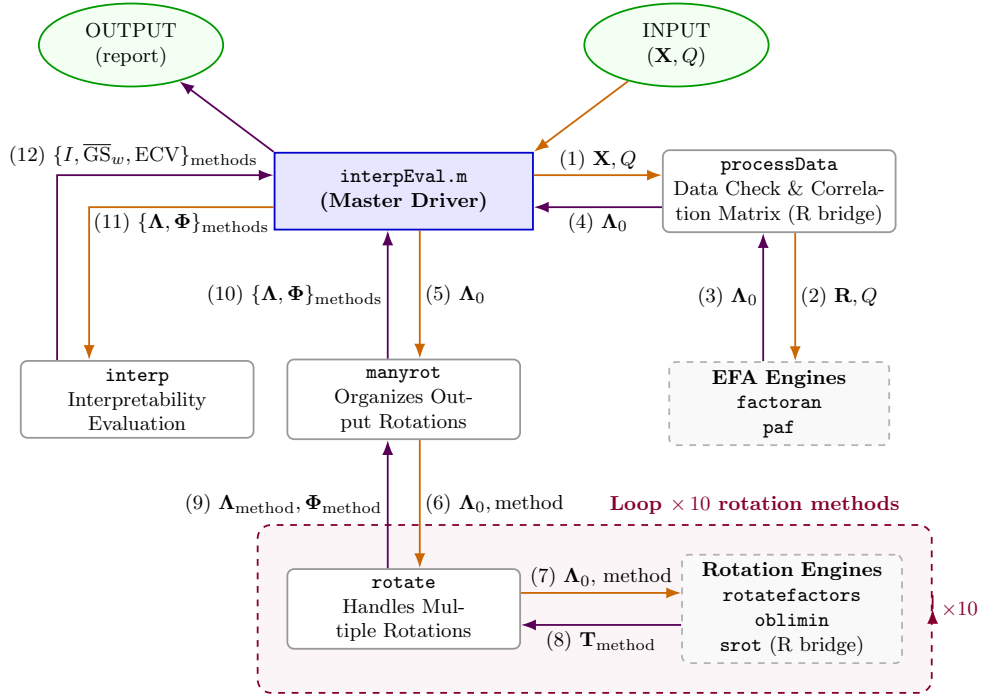
\begin{figure}[h!]
\centering
\resizebox{\textwidth}{!}{%
\begin{tikzpicture}[
    node distance=1.8cm and 2.2cm,
    master/.style={rectangle, draw=blue!80!black, fill=blue!10, thick, text width=3.8cm, align=center, minimum height=1.2cm, font=\bfseries\small},
    func/.style={rectangle, draw=gray!80, fill=white, rounded corners=3pt, thick, text width=3.4cm, align=center, minimum height=1.1cm, font=\small},
    subfunc/.style={rectangle, draw=gray!60, fill=gray!5, rounded corners=3pt, dashed, thick, text width=3.2cm, align=center, minimum height=1cm, font=\small},
    io/.style={ellipse, draw=green!60!black, fill=green!5, thick, text width=1.8cm, align=center, minimum height=0.8cm, font=\small},
    arrow/.style={-Latex, thick, draw=blue!80!black},
    lbl/.style={font=\scriptsize, inner sep=2pt}
]
% --- NODES POSITIONING ---
\node [master] (driver) {\texttt{interpEval.m}\\(Master Driver)};

% Top-Left Input Cloud & Top-Right Output Cloud
\node [io, above left=1.2cm and 1.2cm of driver] (output) {OUTPUT\\ (report)};
\node [io, above right=1.2cm and 1.2cm of driver] (in_data) {INPUT\\$(\mathbf{X}, Q)$};

% Right Functional Block & Sub-module
\node [func, right=2.0cm of driver] (process) {\textbf{\texttt{processData}}\\Data Check \& Correlation Matrix (R bridge)};
\node [subfunc, below=2.0cm of process] (efa_engines) {\textbf{EFA Engines}\\
\texttt{factoran}\\
\texttt{paf}};

% Bottom Row Functional Blocks
\node [func, below=2.0cm of driver] (manyrot) {\textbf{\texttt{manyrot}}\\Organizes Output Rotations};
\node [func, left=0.5cm of manyrot] (interp) {\textbf{\texttt{interp}}\\Interpretability Evaluation};
\node [func, below=2.0cm of manyrot] (rotate) {\textbf{\texttt{rotate}}\\Handles Multiple Rotations};

% Sub-module below rotate
\node [subfunc, right=2.5cm of rotate] (rot_engines) {\textbf{Rotation Engines}\\
\texttt{rotatefactors}\\
\texttt{oblimin}\\
\texttt{srot} (R bridge)};

% --- LOOP GROUPING ---
\begin{pgfonlayer}{background}
\node [draw=purple!70!black, dashed, thick, rounded corners=6pt,
       fill=purple!4, inner sep=0.45cm,
       fit=(rotate)(rot_engines)] (loopbox) {};
\end{pgfonlayer}
\node [font=\scriptsize\bfseries, purple!70!black,
       above=1pt of loopbox.north, xshift=2.5cm] (looplabel)
       {Loop $\times\,10$ rotation methods};
\draw[-Latex, thick, purple!70!black]
    (loopbox.east) to[out=20, in=-20, looseness=6]
    node[right, font=\scriptsize\bfseries, purple!70!black] {$\times 10$}
    (loopbox.east);

% --- ARROWS & DATA FLOW ---
\draw [arrow, draw=orange!80!black] (in_data) -- node[lbl, above left] {} (driver.north east);
\draw [arrow, draw=violet!70!black] (driver.north west) -- node[lbl, above right] {} (output);

\draw [arrow, draw=orange!80!black] ($(driver.east)+(0, 0.25)$) -- node[lbl, above] {(1) $\mathbf{X},Q$} ($(process.west)+(0, 0.25)$);
\draw [arrow, draw=violet!70!black] ($(process.west)-(0, 0.25)$) -- node[lbl, below] {(4) $\mathbf{\Lambda}_0$} ($(driver.east)-(0, 0.25)$);

\draw [arrow, draw=orange!80!black] ($(process.south)+(0.25, 0)$) -- node[lbl, right] {(2) $\mathbf{R}, Q$} ($(efa_engines.north)+(0.25, 0)$);
\draw [arrow, draw=violet!70!black] ($(efa_engines.north)-(0.25, 0)$) -- node[lbl, left] {(3) $\mathbf{\Lambda}_0$} ($(process.south)-(0.25, 0)$);

\draw [arrow, draw=orange!80!black] ($(driver.south)+(0.25, 0)$) -- node[lbl, right] {(5) $\mathbf{\Lambda}_0$} ($(manyrot.north)+(0.25, 0)$);
\draw [arrow, draw=violet!70!black] ($(manyrot.north)-(0.25, 0)$) -- node[lbl, left] {(10) $\{\mathbf{\Lambda},\mathbf{\Phi}\}_{\text{methods}}$} ($(driver.south)-(0.25, 0)$);

\draw [arrow, draw=orange!80!black] ($(manyrot.south)+(0.25, 0)$) -- node[lbl, right] {(6) $\mathbf{\Lambda}_0, \text{method}$} ($(rotate.north)+(0.25, 0)$);
\draw [arrow, draw=violet!70!black] ($(rotate.north)-(0.25, 0)$) -- node[lbl, left] {(9) $\mathbf{\Lambda}_{\text{method}},\mathbf{\Phi}_{\text{method}}$} ($(manyrot.south)-(0.25, 0)$);

\draw [arrow, draw=orange!80!black] ($(rotate.east)+(0, 0.25)$) -- node[lbl, above] {(7) $\mathbf{\Lambda}_0$, method} ($(rot_engines.west)+(0, 0.25)$);
\draw [arrow, draw=violet!70!black] ($(rot_engines.west)-(0, 0.25)$) -- node[lbl, below] {(8) $\mathbf{T}_{\text{method}}$} ($(rotate.east)-(0, 0.25)$);

\draw [arrow, draw=orange!80!black] ($(driver.west)-(0, 0.25)$) -| node[lbl, near start, below] {(11) $\{\mathbf{\Lambda},\mathbf{\Phi}\}_{\text{methods}}$} ($(interp.north)-(0.75, 0)$);
\draw [arrow, draw=violet!70!black] ($(interp.north)-(1.25, 0)$) |- node[lbl, near end, above, xshift = -0.5cm, yshift = 0.05cm] {(12) $\{I,\overline{\operatorname{GS}}_w,\operatorname{ECV}\}_{\text{methods}}$} ($(driver.west)+(0, 0.25)$);

\end{tikzpicture}
}
\caption{Data flow diagram illustrating the nested architecture of the performance evaluation routine \texttt{interpEval.m}.}
\label{fig4}
\end{figure}

Figure \ref{fig4} details the internal data flow of \texttt{interpEval}, the master driver that automates the full model-evaluation pipeline. Arrows are colored to distinguish their role in the pipeline: the orange ones mark inputs, while the violet ones mark outputs.

Step (1) shows \texttt{processData} receiving the raw data matrix $\mathbf{X}$ and the number of factors $Q$. Then, the function checks if ordinal data is detected; if so, data is sent to an external \texttt{R} routine that builds the polychoric correlation matrix (heterogeneous if both continuous and ordinal data is detected) through the \texttt{R} \texttt{psych} package \citep{revelle2015package}. In step (2), \texttt{processData} passes the correlation matrix $\mathbf{R}$ and $Q$ to the EFA Engines (MATLAB's \texttt{factoran} or our own \texttt{paf} routine), which apply Maximum Likelihood or Principal Axis Factoring based on the data's nature. In step (3), the EFA Engines return the unrotated pattern matrix $\mathbf{\Lambda}_0$ to \texttt{processData}, which concludes data diagnostics and passes $\mathbf{\Lambda}_0$ to the Master driver in step (4). In step (5), the unrotated pattern matrix is passed to \texttt{manyrot}. The dashed box highlights the core iterative rotation block: for each of the ten candidate rotation criteria, \texttt{manyrot} calls \texttt{rotate} with $\mathbf{\Lambda}_0$ and the method name (step 6); \texttt{rotate} in turn invokes the appropriate rotation engine (MATLAB's \texttt{rotatefactors} or our own \texttt{oblimin} or \texttt{srot} routines) to obtain the corresponding rotation matrix $\mathbf{T}_{\text{method}}$ (steps 7--8). From $\mathbf{T}_{\text{method}}$, \texttt{rotate} computes the rotated loading matrix $\mathbf{\Lambda}_{\text{method}} = \mathbf{\Lambda}_0 \mathbf{T}_{\text{method}}$ and the associated factor correlation matrix $\mathbf{\Phi}_{\text{method}} = (\mathbf{T}_{\text{method}}^\top \mathbf{T}_{\text{method}})^{-1}$, returning the pair to \texttt{manyrot} (step 9). This loop is executed ten times, once per rotation method, and \texttt{manyrot} collects the full set $\{\mathbf{\Lambda}, \mathbf{\Phi}\}_{\text{methods}}$ before forwarding it back to the driver (step 10). In step (11), the driver sends these results to \texttt{interp}, which evaluates, for every candidate rotation, the Interpretability Index $I$ together with its two constituents $\operatorname{ECV}$ and $\overline{\mathrm{GS}}_w$ (step 12), and \texttt{interpEval} assembles these results into the final report handed back to the user. As an additional diagnostic step, each $\mathbf{\Lambda}_{\text{method}}$ is passed to \texttt{rowcompliance}, which returns the compliance scores $F(\mathbf{\Lambda}_{\text{method}})$. Whenever two or more solutions lie within a user-defined tolerance $\epsilon$ of the maximum $I$ (default $\epsilon = 5\%$), the tie is broken by selecting the rotation with the lowest macro-average failure rate $F(\mathbf{\Lambda})$.

% =========================================================================
\section{Structural Compliance Indices: A Row-Wise Complement to the Interpretability Index}
\label{sec:thurstone}
% =========================================================================

The interpretability index aggregates information at the model level, meaning that competing rotations can attain interpretability scores that differ only marginally (often at the third decimal place).

In such cases, $I$ alone cannot definitively determine the optimal solution. To address this uncertainty, we propose a complementary row-wise diagnostic tool that operationalizes classical qualitative principles of simple structure \citep{thurstone1947} into threshold-based quantitative checks, building on the row-wise complexity index of \cite{Hofmann1978} and best-practice threshold guidelines \citep{CostelloOsborne2005}.

Let $\mathbf{\Lambda} \in \mathbb{R}^{J \times Q}$ be a rotated pattern matrix and $0 < \tau_0 < \tau_1$ define thresholds for negligible and salient loadings respectively (defaulting to $\tau_0 = 0.10$ and $\tau_1 = 0.40$ as per common practice in the EFA literature). For each factor $q = 1,\dots,Q$ we define the sets of negligible ($\mathcal{N}_q$) and salient ($\mathcal{S}_q$) variable indices as:
\begin{equation}
    \mathcal{N}_q = \{ j \in \{1,\dots,J\} : |\lambda_{jq}| \le \tau_0 \}; \qquad 
    \mathcal{S}_q = \{ j \in \{1,\dots,J\} : |\lambda_{jq}| \ge \tau_1 \}.
\end{equation}

The classical qualitative principles of simple structure are operationalized here into five diagnostic checks:

\begin{description}
\item[Query 1 -- Row Simplicity.] Every variable must have a negligible loading on at least one factor:
\begin{equation}
    \sum_{q=1}^{Q} \mathbb{I}(j \in \mathcal{N}_q) \ge 1, \qquad \forall j = 1,\dots,J.
    \label{eq:q1}
\end{equation}

\item[Query 2 -- Minimum Negligible Loadings.] Every factor must contain at least $Q$ negligible loadings:
\begin{equation}
    \operatorname{card}(\mathcal{N}_q) \ge Q, \qquad \forall q = 1,\dots,Q.
    \label{eq:q2}
\end{equation}

\item[Query 3 -- Factor Differentiation.] For every ordered pair of distinct factors $(q,r)$, at least half of the variables salient on $q$ must be negligible on $r$:
\begin{equation}
    P_{q\rightarrow r} = \frac{\operatorname{card}(\mathcal{S}_q \cap \mathcal{N}_r)}{\operatorname{card}(\mathcal{S}_q)} \ge \frac{1}{2}, \qquad \text{provided } \operatorname{card}(\mathcal{S}_q) > 0.
    \label{eq:q3}
\end{equation}
If $\operatorname{card}(\mathcal{S}_q) = 0$, the ratio is undefined and flagged as an automatic diagnostic failure. The check is required symmetrically ($P_{r\rightarrow q} \ge \frac{1}{2}$).

\item[Query 4 -- Shared Independence.] Every pair of distinct factors $(q,r)$ must share a common set of inactive variables. Specifically, the Jaccard similarity between their sets of negligible loadings must be at least $1/2$, meaning at least half of the variables that are negligible on either factor must be negligible on both:
\begin{equation}
    J^{(N)}_{qr} = \frac{\operatorname{card}(\mathcal{N}_q \cap \mathcal{N}_r)}{\operatorname{card}(\mathcal{N}_q \cup \mathcal{N}_r)} \ge \frac{1}{2}.
    \label{eq:q4}
\end{equation}
If $\operatorname{card}(\mathcal{N}_q \cup \mathcal{N}_r) = 0$, the ratio is undefined and flagged as a diagnostic failure.

\item[Query 5 -- Low Cross-Loading Complexity.] For every pair of distinct factors $(q,r)$, cross-loading variables should remain minimal:
\begin{equation}
    \operatorname{card}(\mathcal{S}_q \cap \mathcal{S}_r) \le \frac{1}{Q+1} \sum_{k=1}^{Q} \operatorname{card}(\mathcal{S}_k).
    \label{eq:q5}
\end{equation}
\end{description}

Where the scaling factor $1/(Q+1)$ prevents the check from becoming trivially easy to pass in higher dimensions. To summarize compliance, let $f_k$ denote the number of failed checks for criterion $k$, and $n_k$ denote the total valid checks conducted for that criterion. The criterion failure rate percentage is $F_k = 100 \times (f_k / n_k)$ for all $k=1,\dots,5$.

The overall compliance score is defined as:
\begin{equation}
    F(\mathbf{\Lambda}) = \frac{1}{5} \sum_{k=1}^{5} F_k,
    \label{eq:thurstone-macro}
\end{equation}
where $F(\mathbf{\Lambda}) = 0\%$ indicates full compliance and higher values signal increasing deviations from simple structure.

In practice, when two candidate rotations $s_1$ and $s_2$ yield nearly identical interpretability scores (e.g., $(1-\epsilon) I_{s_2} \le I_{s_1} \le I_{s_2}$), $F(\mathbf{\Lambda})$ acts as a secondary tie-breaker, favoring the solution with the lower failure rate. Convergence across both high $I$ scores and low $F(\mathbf{\Lambda})$ values confirms that the underlying factor structure reflects real data properties rather than artifacts of a specific rotation criterion. The tolerance parameter ($\epsilon$) defines the maximum proportional loss in the interpretability index that is considered practically negligible.

% =========================================================================
\section{Applications}
\label{sec:applications}
% =========================================================================

In this section, the procedure described in Figure \ref{fig4} will be applied to five established benchmark datasets from the statistical literature, spanning diverse sample sizes, variable counts, and distributional properties:
\begin{itemize}
    \item Wine \citep{wine_109} $\rightarrow \{N = 178, J = 13, Q = 3\}$
    \item Psychological Test \citep{HolzingerSwineford1939} $\rightarrow \{N = 301, J = 24, Q = 4\}$
    \item Breast Cancer Wisconsin \citep{breastcancer} $\rightarrow \{N = 569, J = 10, Q = 2\}$
    \item Boston Housing \citep{HarrisonRubinfeld1978} $\rightarrow \{N = 506, J = 13, Q = 3\}$
    \item International Personality \citep{revelle2010individual} $\rightarrow \{N = 2794, J = 26, Q = 5\}$
\end{itemize}

For each dataset, the number of retained factors $Q$ is selected from the original paper's analysis where available, or according to the procedure in Section \ref{sec:choice_q} otherwise. As anticipated in Section \ref{sec:workflow_software}, the unrotated loading matrix $\mathbf{\Lambda}_0$ is subject to ten analytic rotation methods, covering: orthogonal algorithms such as Equamax \citep{Saunders1962}, Parsimax \citep{CrawfordFerguson1970}, Quartimax \citep{NeuhausWrigley1954}, and Varimax \citep{kaiser1958}; standard oblique transformations like Oblimin \citep{JennrichSampson1966} and Promax \citep{HendricksonWhite1964}; sparse techniques like Bentler \citep{bentler1977factor}, Geomin \citep{Yates1987}, Infomax \citep{McKeon1968}, and McCammon \citep{McCammon1966}.

For every dataset, we apply the two-stage evaluation criterion introduced in Sections \ref{sec:workflow_software} and \ref{sec:thurstone}. First, the near-optimal group (which we denote $\mathcal{G}_{1-\epsilon}$) is identified among the eleven candidate solutions (unrotated plus ten rotations) reported in Table \ref{tab:61}. Since $\operatorname{ECV}$ is constant across rotations for a given $Q$, membership in $\mathcal{G}_{1-\epsilon}$ is driven entirely by $\overline{\mathrm{GS}}_w$. Second, row-wise compliance is computed for every member of $\mathcal{G}_{1-\epsilon}$, and the rotation with the lowest $F(\mathbf{\Lambda})$ within the group is reported as the final recommendation. We set $\epsilon = 0.05$ a priori, so that only rotations attaining at least 95\% of the maximum observed index value are eligible for secondary assessment. The value 0.05 should be understood as an operational tolerance rather than as a universal theoretical threshold.

\begin{longtable}{lccc}
\caption{Interpretability and Structural Evaluation Across Benchmark Datasets. Rotations in $\mathcal{G}_{0.95}$ are underlined.} \label{tab:61} \\
\small
\renewcommand{\arraystretch}{0.6}
\setlength{\tabcolsep}{12pt} \\
\toprule
\textbf{Method} & $I$ & $\overline{\mathrm{GS}}_w$ & \textbf{Orth} \\
\midrule
\endfirsthead

\toprule
\textbf{Method} & $I$ & $\overline{\mathrm{GS}}_w$ & \textbf{Orth} \\
\midrule
\endhead

\multicolumn{4}{r}{\small\itshape continued on next page\ldots} \\
\endfoot

\bottomrule
\endlastfoot

\multicolumn{4}{c}{\textbf{Wine Data} ($\operatorname{ECV} = 0.584$)} \\
\midrule
UNROTATED & 0.442 & 0.334 & 1.000 \\
equamax   & 0.490 & 0.411 & 1.000 \\
parsimax  & 0.490 & 0.412 & 1.000 \\
quartimax & 0.485 & 0.403 & 1.000 \\
varimax   & 0.488 & 0.408 & 1.000 \\
\underline{promax}    & 0.520 & 0.463 & 0.743 \\
\underline{oblimin}   & 0.504 & 0.435 & 0.830 \\
\underline{bentler}   & 0.508 & 0.443 & 0.766 \\
geomin    & 0.475 & 0.386 & 0.818 \\
\underline{infomax}   & 0.506 & 0.439 & 0.651 \\
mccammon  & 0.478 & 0.391 & 1.000 \\
\midrule
\multicolumn{4}{c}{\textbf{Holzinger--Swineford Psychological Tests} ($\operatorname{ECV} = 0.460$)} \\
\midrule
UNROTATED & 0.326 & 0.231 & 1.000 \\
equamax   & 0.439 & 0.420 & 1.000 \\
parsimax  & 0.439 & 0.419 & 1.000 \\
quartimax & 0.437 & 0.416 & 1.000 \\
varimax   & 0.440 & 0.422 & 1.000 \\
\underline{promax}    & 0.488 & 0.519 & 0.521 \\
\underline{oblimin}   & 0.483 & 0.507 & 0.688 \\
\underline{bentler}   & 0.488 & 0.517 & 0.585 \\
\underline{geomin}    & 0.484 & 0.510 & 0.650 \\
\underline{infomax}   & 0.489 & 0.520 & 0.558 \\
mccammon  & 0.440 & 0.422 & 1.000 \\
\midrule
\multicolumn{4}{c}{\textbf{Breast Cancer Wisconsin Data} ($\operatorname{ECV} = 0.767$)} \\
\midrule
UNROTATED & 0.415 & 0.225 & 1.000 \\
equamax   & 0.481 & 0.302 & 1.000 \\
parsimax  & 0.481 & 0.302 & 1.000 \\
quartimax & 0.481 & 0.301 & 1.000 \\
varimax   & 0.481 & 0.302 & 1.000 \\
\underline{promax}    & 0.508 & 0.337 & 0.663 \\
\underline{oblimin}   & 0.500 & 0.326 & 0.778 \\
\underline{bentler}   & 0.504 & 0.331 & 0.730 \\
\underline{geomin}    & 0.502 & 0.329 & 0.747 \\
\underline{infomax}   & 0.504 & 0.331 & 0.719 \\
mccammon  & 0.482 & 0.302 & 1.000 \\
\midrule
\multicolumn{4}{c}{\textbf{Boston Housing Data} ($\operatorname{ECV} = 0.668$)} \\
\midrule
UNROTATED & 0.321 & 0.154 & 1.000 \\
equamax   & 0.511 & 0.391 & 1.000 \\
parsimax  & 0.511 & 0.391 & 1.000 \\
quartimax & 0.452 & 0.306 & 1.000 \\
varimax   & 0.511 & 0.390 & 1.000 \\
\underline{promax}    & 0.601 & 0.541 & 0.395 \\
\underline{oblimin}   & 0.592 & 0.525 & 0.523 \\
\underline{bentler}   & 0.598 & 0.535 & 0.439 \\
\underline{geomin}    & 0.589 & 0.519 & 0.529 \\
\underline{infomax}   & 0.600 & 0.538 & 0.413 \\
mccammon  & 0.509 & 0.388 & 1.000 \\
\midrule
\multicolumn{4}{c}{\textbf{International Personality Item Pool Data} ($\operatorname{ECV} = 0.457$)} \\
\midrule
UNROTATED & 0.333 & 0.243 & 1.000 \\
\underline{equamax}   & 0.478 & 0.499 & 1.000 \\
\underline{parsimax}  & 0.478 & 0.499 & 1.000 \\
\underline{quartimax} & 0.476 & 0.496 & 1.000 \\
\underline{varimax}   & 0.477 & 0.498 & 1.000 \\
\underline{promax}    & 0.496 & 0.537 & 0.763 \\
\underline{oblimin}   & 0.490 & 0.526 & 0.833 \\
\underline{bentler}   & 0.491 & 0.528 & 0.753 \\
\underline{geomin}    & 0.477 & 0.497 & 0.872 \\
\underline{infomax}   & 0.488 & 0.522 & 0.814 \\
\underline{mccammon}  & 0.474 & 0.491 & 1.000 \\
\end{longtable}

We begin with the Wine dataset. The unrotated solution yields a low sparseness score ($\overline{\mathrm{GS}}_w = 0.334$), resulting in a modest $I = 0.442$ due to heavy cross-loadings across variables. Applying standard orthogonal rotations prevents factor correlation but elevates sparseness into the $0.403\text{--}0.412$ range, with Equamax and Parsimax achieving the highest Interpretability Index ($I = 0.490$).

The highest scores are achieved by standard oblique transformations. Promax attains the maximum Generalized Sparseness ($\overline{\mathrm{GS}}_w = 0.463$) and hence Interpretability Index ($I = 0.520$), closely followed by Bentler ($I = 0.508,\ \overline{\mathrm{GS}}_w = 0.443$). A significant departure from strict orthogonality (e.g., $\mathrm{Orth}(\mathbf{\Phi}) = 0.743$ for Promax) indicates a non-negligible root-mean-square correlation among the extracted factors, allowing the model more freedom to allocate loadings across correlated dimensions, which in turn enables the latent factors to align more naturally with distinct chemical profiles.

Applying the grouping criterion described in Section \ref{sec:thurstone}, we have $\mathcal{G}_{0.95}^{\text{Wine}} = \{\text{Promax}, \text{Bentler}, \text{Infomax}, \text{Oblimin}\}$ (underlined in Table \ref{tab:61}). Row-wise compliance for $\mathcal{G}_{0.95}^{\text{Wine}}$ is reported in Table \ref{tab:wine-rowcompliance}.

\begin{table}[h]
\centering
\caption{Percentage of failed elementary checks per query and overall compliance score, Wine dataset, restricted to $\mathcal{G}_{0.95}$.}
\label{tab:wine-rowcompliance}
\begin{tabular}{lccccc c}
\toprule
\textbf{Method} & \textit{Q1} & \textit{Q2} & \textit{Q3} & \textit{Q4} & \textit{Q5} & $F$ (Total) \\
\midrule
promax  & 23.08 & 66.67 & 66.67  & 100.00 & 0.00 & 51.28 \\
oblimin & 69.23 & 66.67 & 100.00 & 100.00 & 0.00 & 67.18 \\
bentler & 46.15 & 66.67 & 100.00 & 100.00 & 0.00 & 62.56 \\
infomax & 30.77 & 66.67 & 66.67  & 100.00 & 0.00 & 52.82 \\
\bottomrule
\end{tabular}
\end{table}

Within $\mathcal{G}_{0.95}^{\text{Wine}}$, Promax attains the lowest overall failure rate ($F = 51.28\%$), and is therefore the recommended solution. This improves on the classical Varimax rotation historically used for exploratory analyses of the Wine chemical panel ($I=0.488$ and hence excluded from $\mathcal{G}_{0.95}$). This is consistent with the presence of correlated underlying chemical characteristics and illustrates how allowing oblique factors can improve the proposed structural-simplicity criterion

For the Holzinger--Swineford data, $\mathcal{G}_{0.95}$ consists of \{Infomax, Promax, Bentler, Geomin, Oblimin\}, all scoring $I \in [0.483, 0.489]$.

\begin{table}[ht!]
\centering
\caption{Percentage of failed elementary checks per query and overall compliance score, Holzinger--Swineford Psychological Tests data, restricted to $\mathcal{G}_{0.95}$.}
\label{tab:thurstone-psy}
\begin{tabular}{lccccc c}
\toprule
\textbf{Method} & \textit{Q1} & \textit{Q2} & \textit{Q3} & \textit{Q4} & \textit{Q5} & $F$ (Total) \\
\midrule
promax  & 8.33 & 0.00 & 50.00 & 16.67 & 0.00 & 15.00 \\
oblimin & 4.17 & 0.00 & 50.00 & 0.00  & 0.00 & 10.83 \\
bentler & 4.17 & 0.00 & 50.00 & 16.67 & 0.00 & 14.17 \\
geomin  & 0.00 & 0.00 & 16.67 & 0.00  & 0.00 & 3.33  \\
infomax & 0.00 & 0.00 & 33.33 & 0.00  & 0.00 & 6.67  \\
\bottomrule
\end{tabular}
\end{table}

Table \ref{tab:thurstone-psy} reports the row-wise compliance scores for this group. Although Infomax attains the nominal maximum ($I=0.489$), it is Geomin, the second lowest-$I$ member of $\mathcal{G}_{0.95}$, that achieves the best row-wise compliance ($F=3.33\%$ against $6.67\%$ for Infomax, $10.83\%$ for Oblimin, $14.17\%$ for Bentler, and $15.00\%$ for Promax). Geomin is therefore the recommended solution. This improves on the classical Varimax rotation ($I=0.440$, excluded from $\mathcal{G}_{0.95}$), and is consistent with the original group-factor interpretation of \cite{HolzingerSwineford1939}. In fact, the 24 mental tests load on four correlated, well-separated group factors (spatial, verbal, speed, and memory), a structure that an orthogonal rotation might blur.

Table \ref{tab:factor_loadings} shows the pattern matrices of the (a) Geomin and (b) Infomax solutions of the Psychological Test data. Clearly, the inspection of large loading matrices is problematic, and the complexity scales when more solutions need comparison.

\begin{table}[htbp]
  \centering 
  \footnotesize
  % --- Subtable 1: Geomin ---
  \begin{subtable}[t]{0.48\textwidth}
    \centering
    \caption{$\mathbf{\Lambda}^{(\text{geomin})}$}
    \label{tab3a}
    \begin{tabular}{crrrr}
      \toprule
      Item & \multicolumn{1}{c}{F1} & \multicolumn{1}{c}{F2} & \multicolumn{1}{c}{F3} & \multicolumn{1}{c}{F4} \\
      \midrule
      1  &  0.093 & -0.632 &  0.009 & -0.022 \\
      2  & -0.025 & -0.588 &  0.058 &  0.128 \\
      3  &  0.050 & -0.460 &  0.121 & -0.022 \\
      4  & -0.178 & -0.691 & -0.045 & -0.030 \\
      5  &  0.856 &  0.025 &  0.126 & -0.054 \\
      6  &  0.807 & -0.015 & -0.028 & -0.007 \\
      7  &  0.924 &  0.049 &  0.091 &  0.019 \\
      8  &  0.691 & -0.089 & -0.002 & -0.024 \\
      9  &  0.816 & -0.097 &  0.021 &  0.031 \\
      10 &  0.033 &  0.278 & -0.006 & -0.862 \\
      11 &  0.187 &  0.016 & -0.114 & -0.550 \\
      12 & -0.045 & -0.100 &  0.121 & -0.671 \\
      13 & -0.013 & -0.310 &  0.084 & -0.485 \\
      14 &  0.091 &  0.046 & -0.666 &  0.038 \\
      15 & -0.141 & -0.114 & -0.572 &  0.008 \\
      16 &  0.003 & -0.356 & -0.420 & -0.025 \\
      17 & -0.017 &  0.139 & -0.503 & -0.318 \\
      18 &  0.014 & -0.071 & -0.374 & -0.186 \\
      19 &  0.159 & -0.141 & -0.293 & -0.053 \\
      20 &  0.194 & -0.458 & -0.161 &  0.087 \\
      21 &  0.128 & -0.352 & -0.038 & -0.317 \\
      22 &  0.376 & -0.393 & -0.033 &  0.017 \\
      23 &  0.240 & -0.520 & -0.048 & -0.088 \\
      24 &  0.336 & -0.086 & -0.142 & -0.321 \\
      \bottomrule
    \end{tabular}
  \end{subtable}
  \hfill
  % --- Subtable 2: Infomax ---
  \begin{subtable}[t]{0.48\textwidth}
    \centering
    \caption{$\mathbf{\Lambda}^{(\text{infomax})}$}
    \label{tab3b}
    \begin{tabular}{crrrr}
      \toprule
      Item & \multicolumn{1}{c}{F1} & \multicolumn{1}{c}{F2} & \multicolumn{1}{c}{F3} & \multicolumn{1}{c}{F4} \\
      \midrule
      1  &  0.020 &  0.612 &  0.050 &  0.073 \\
      2  & -0.038 &  0.592 & -0.102 & -0.035 \\
      3  & -0.113 &  0.461 &  0.057 &  0.039 \\
      4  &  0.081 &  0.670 &  0.060 & -0.209 \\
      5  & -0.134 & -0.040 &  0.057 &  0.875 \\
      6  &  0.039 & -0.016 & -0.008 &  0.820 \\
      7  & -0.096 & -0.064 & -0.026 &  0.948 \\
      8  &  0.012 &  0.063 &  0.018 &  0.699 \\
      9  & -0.012 &  0.074 & -0.039 &  0.830 \\
      10 & -0.008 & -0.345 &  0.897 & -0.007 \\
      11 &  0.127 & -0.083 &  0.564 &  0.154 \\
      12 & -0.132 &  0.058 &  0.727 & -0.081 \\
      13 & -0.080 &  0.273 &  0.536 & -0.047 \\
      14 &  0.740 & -0.134 & -0.121 &  0.065 \\
      15 &  0.641 &  0.039 & -0.068 & -0.174 \\
      16 &  0.485 &  0.288 & -0.007 & -0.031 \\
      17 &  0.553 & -0.229 &  0.271 & -0.054 \\
      18 &  0.420 &  0.003 &  0.155 & -0.017 \\
      19 &  0.334 &  0.088 &  0.026 &  0.140 \\
      20 &  0.203 &  0.425 & -0.094 &  0.179 \\
      21 &  0.059 &  0.307 &  0.344 &  0.098 \\
      22 &  0.058 &  0.368 & -0.011 &  0.369 \\
      23 &  0.080 &  0.485 &  0.106 &  0.219 \\
      24 &  0.165 &  0.027 &  0.321 &  0.314 \\
      \bottomrule
    \end{tabular}
  \end{subtable}
  \caption{Side-by-Side Comparison of Geomin and Infomax Pattern Matrices for the Holzinger--Swineford Psychological Tests data.}
  \label{tab:factor_loadings} 
\end{table}

The \texttt{rowcompliance} routine provides a concise diagnostic summary with a standardized output that indicates to the user where to look for violations of simple structure. Figure \ref{fig:console_output} shows the output of this routine when applied to the Infomax-rotated pattern matrix $\mathbf{\Lambda}^{(\text{infomax})}$ for the data at hand.
At the end of the output, an overall failure percentage of $6.67\%$ indicates a strong alignment with simple structure. This output indicates exactly where the matrix is ``badly conditioned''. For instance, \textit{Query 3} was failed for two distinct pairs of factors. Let's check the pair $(1,2)$: the query requires the majority of the salient loadings of factor 1 to be negligible loadings in factor 2 (and vice versa). Table \ref{tab3b} confirms the output's count: factor 1 has five salient loadings ($0.740, 0.641, 0.485, 0.553, 0.420$), while the corresponding rows of factor 2 only have two negligible loadings ($0.039$ at row 15 and $0.003$ at row 18) out of 3 required.
\begin{figure}[ht!]
\centering
\begin{minipage}{\linewidth}
\begin{lstlisting}[style=matlabconsole]
>> rowcompliance(Lambda_info)
========================================================================
            ROTATION SIMPLICITY COMPLIANCE REPORT (J = 24, Q = 4)
========================================================================

------------------------------------------------------------------------
QUERY 1: ROW SIMPLICITY
Description: Each row j (variable) must contain at least one negligible
             loading (|lambda(j,.)| <= 0.10) across all factors.
------------------------------------------------------------------------
-> PASSED: All 24 rows contain at least one negligible loading.

------------------------------------------------------------------------
QUERY 2: NEGLIGIBLE-LOADING REQUIREMENT
Description: Each column q (factor) must contain at least Q negligible loadings
             (|lambda(.,q)| <= 0.10) across all variables.
------------------------------------------------------------------------
-> PASSED: All 4 factor columns satisfy the overdetermination threshold.

------------------------------------------------------------------------
QUERY 3: FACTOR DIFFERENTIATION
Description: For every pair of factors (q,r), a majority (>= 50%) of the
             variables that are salient (|lambda| >= 0.40) on factor q must
             have negligible loadings (|lambda| <= 0.10) on factor r, AND
             vice versa (r -> q).
------------------------------------------------------------------------
couple (1,2) [Dir: 1->2] FAILED -> Salient on 1: 5; Corresponding zeros on 2: 2 (Required majority: 3)
couple (1,3) [Dir: 1->3] FAILED -> Salient on 1: 5; Corresponding zeros on 3: 2 (Required majority: 3)

------------------------------------------------------------------------
QUERY 4: SHARED INDEPENDENCE
Description: For every pair of factors (q,r), the majority of negligible loadings
             (|lambda| <= 0.10) must be shared.
------------------------------------------------------------------------
-> PASSED: All 6 factor pairs satisfy the shared zero threshold.

------------------------------------------------------------------------
QUERY 5: LOW COMPLEXITY
Description: For every pair of factors (q,r), only a minimal number of variables
             may be salient (|lambda| >= 0.40) on both factors at once.
             Threshold: <= 4 items (scaled inversely by 1/(Q+1) = 20.0%
             across all 20 salient variables in the matrix).
------------------------------------------------------------------------
-> PASSED: All 6 factor pairs satisfy minimal cross-loading limits.

========================================================================
                        EXPLICIT CHECK BREAKDOWN                         
========================================================================
 Query 1 (Row Simplicity)         :  0 / 24 failed (0.00%)
 Query 2 (Negligible Requirement) :  0 /  4 failed (0.00%)
 Query 3 (Factor Differentiation) :  2 /  6 failed (33.33%)
 Query 4 (Shared Independence)    :  0 /  6 failed (0.00%)
 Query 5 (Low Complexity)         :  0 /  6 failed (0.00%)
------------------------------------------------------------------------
 TOTAL CHECKS EVALUATED          : 46
 TOTAL CHECKS FAILED              : 2
 MEAN FAILURE PERCENTAGE (UNWTD) : 6.67%
 COMPLIANCE SCORE                 : 0.9333
========================================================================
\end{lstlisting}
\end{minipage}
\caption{Raw MATLAB Console Output for the \texttt{rowcompliance} routine on the Infomax-rotated solution of Holzinger--Swineford Psychological Tests data.}
\label{fig:console_output}
\end{figure}

Given the size of the routine, we will not include the Geomin output. However, we will use it to flag the only critical instance on $\mathbf{\Lambda}^{(\text{geomin})}$. \textit{Query 3} is failed by the pair (3,2): the former has 4 salient loadings ($-0.666, -0.572, -0.420, -0.503$) at rows 14--17, while the latter's corresponding rows only have a single negligible loading (row 14, $0.046$).

For the Breast Cancer Wisconsin data, $\mathcal{G}_{0.95}$ consists of \{Promax, Oblimin, Bentler, Geomin, Infomax\}. Row-wise compliance within the group is reported in Table \ref{tab:bc-rowcompliance}.

\begin{table}[h]
\centering
\caption{Percentage of failed elementary checks per query and overall compliance score, Breast Cancer Wisconsin data, restricted to $\mathcal{G}_{0.95}$.}
\label{tab:bc-rowcompliance}
\begin{tabular}{lccccc c}
\toprule
\textbf{Method} & \textit{Q1} & \textit{Q2} & \textit{Q3} & \textit{Q4} & \textit{Q5} & $F$ (Total) \\
\midrule
promax  & 50.00 & 0.00 & 100.00 & 100.00 & 0.00 & 50.00 \\
oblimin & 40.00 & 0.00 & 100.00 & 100.00 & 0.00 & 48.00 \\
bentler & 50.00 & 0.00 & 100.00 & 100.00 & 0.00 & 50.00 \\
geomin  & 40.00 & 0.00 & 100.00 & 100.00 & 0.00 & 48.00 \\
infomax & 40.00 & 0.00 & 100.00 & 100.00 & 0.00 & 48.00 \\
\bottomrule
\end{tabular}
\end{table}

Oblimin, Geomin, and Infomax achieve the lowest failure rate (\(F=48\%\)) within \(G_{0.95}\). Their interpretability scores also exceed that of the excluded Varimax solution (\(I=0.481\)). This is consistent with the well-documented bidimensional structure of the Wisconsin diagnostic features: a ``size'' factor (radius, perimeter, area) and a ``shape/texture'' factor (concavity, symmetry, fractal dimension) \citep{street1993nuclear}, which are known to be moderately correlated in malignant samples, favoring oblique over orthogonal rotations.

For the Boston Housing data, $\mathcal{G}_{0.95}$ again isolates the five oblique/sparse-oblique rotations. Table \ref{tab:boston-rowcompliance} reports the corresponding row-wise diagnostics.

\begin{table}[h]
\centering
\caption{Percentage of failed elementary checks per query and overall compliance score, Boston Housing data, restricted to $\mathcal{G}_{0.95}$.}
\label{tab:boston-rowcompliance}
\begin{tabular}{lccccc c}
\toprule
\textbf{Method} & \textit{Q1} & \textit{Q2} & \textit{Q3} & \textit{Q4} & \textit{Q5} & $F$ (Total) \\
\midrule
promax  & 15.38 & 0.00  & 66.67  & 66.67 & 0.00 & 29.74 \\
oblimin & 23.08 & 33.33 & 100.00 & 66.67 & 0.00 & 44.62 \\
bentler & 23.08 & 0.00  & 100.00 & 66.67 & 0.00 & 37.95 \\
geomin  & 23.08 & 0.00  & 66.67  & 33.33 & 0.00 & 24.62 \\
infomax & 15.38 & 0.00  & 66.67  & 66.67 & 0.00 & 29.74 \\
\bottomrule
\end{tabular}
\end{table}

Because the five candidates in $\mathcal{G}_{0.95}$ are separated by less than 0.02 in $I$, row-wise compliance is decisive here: Geomin attains $F=24.62\%$, yielding the lowest failure rate under the proposed row-compliance criteria. This improves on the classical orthogonal solution (Varimax, $I=0.511$, excluded from $\mathcal{G}_{0.95}$), and is consistent with \cite{HarrisonRubinfeld1978} original hedonic decomposition of housing prices into correlated accessibility/pollution and socioeconomic dimensions, which an orthogonal rotation cannot represent faithfully.

\begin{table}[h]
\centering
\caption{Percentage of failed elementary checks per query and overall compliance score, International Personality Item Pool data, restricted to $\mathcal{G}_{0.95}$.}
\label{tab:ipip-rowcompliance}
\begin{tabular}{lccccc c}
\toprule
\textbf{Method} & \textit{Q1} & \textit{Q2} & \textit{Q3} & \textit{Q4} & \textit{Q5} & $F$ (Total) \\
\midrule
equamax   & 0.00 & 0.00 & 50.00 & 0.00  & 0.00 & 10.00 \\
parsimax  & 0.00 & 0.00 & 50.00 & 0.00  & 0.00 & 10.00 \\
quartimax & 0.00 & 0.00 & 50.00 & 0.00  & 0.00 & 10.00 \\
varimax   & 0.00 & 0.00 & 60.00 & 0.00  & 0.00 & 12.00 \\
promax    & 0.00 & 0.00 & 60.00 & 10.00 & 0.00 & 14.00 \\
oblimin   & 0.00 & 0.00 & 50.00 & 0.00  & 0.00 & 10.00 \\
bentler   & 0.00 & 0.00 & 50.00 & 0.00  & 0.00 & 10.00 \\
geomin    & 0.00 & 0.00 & 30.00 & 0.00  & 0.00 & 6.00  \\
infomax   & 0.00 & 0.00 & 50.00 & 0.00  & 0.00 & 10.00 \\
mccammon  & 0.00 & 0.00 & 40.00 & 0.00  & 0.00 & 8.00  \\
\bottomrule
\end{tabular}
\end{table}

The International Personality Item Pool data exhibit a different pattern: it excludes only the unrotated solution. Unlike the previous four datasets, orthogonal rotations are not penalized here, reflecting the well-documented approximate orthogonality of the Big Five personality dimensions \citep{revelle2010individual}: correlations among the extracted factors remain modest even under oblique rotation ($\mathrm{Orth} \ge 0.753$).

In Table \ref{tab:ipip-rowcompliance}, Geomin achieves the lowest failure rate within the group ($F = 6.00\%$), outperforming both traditional orthogonal staples (such as Quartimax at $10.00\%$ and Varimax at $12.00\%$) and oblique approaches like Promax ($14.00\%$).

Crucially, no rotation triggered \textit{Query 5} in any dataset. This is expected, since the check requires no more than a fixed quota of variables (see Equation \ref{eq:q5}) to be simultaneously salient on two distinct factors. This may happen in case of an unrotated solution or, more likely, when the number of factors $Q$ is misspecified.

% -------------------------------------------------------------------------
\subsection{Choice of \texorpdfstring{$Q$}{Q}}
\label{sec:choice_q}
% -------------------------------------------------------------------------

\begin{figure}[ht!]
    \centering
    \includegraphics[width=0.95\textwidth]{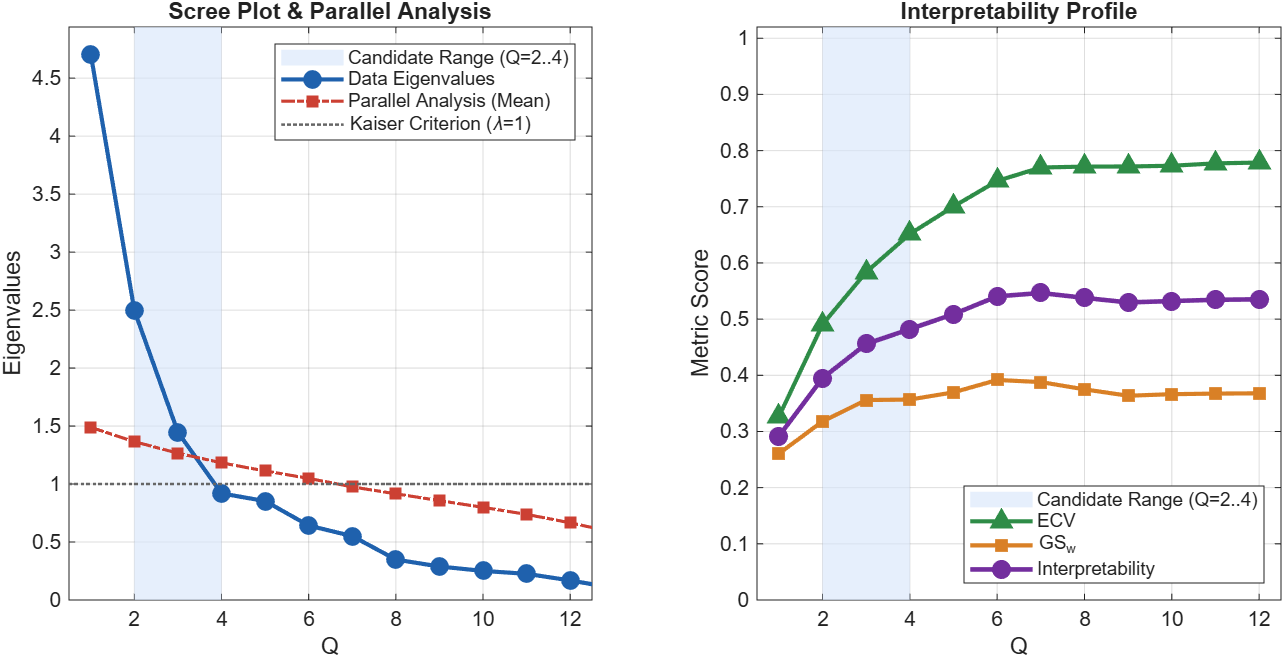}
    \caption{Selection procedure for the number of factors in the Wine dataset applied to the unrotated EFA solution.}
    \label{fig:wine_chooseQ}
\end{figure}

The Interpretability Index also proves to be a valuable tool to determine the optimal number of factors $Q$. The Wine dataset will be used as a case study. Figure \ref{fig:wine_chooseQ} (left) displays standard heuristics used for factor selection, including the classical scree plot of eigenvalues (blue line), the average bootstrapped eigenvalues from parallel analysis \citep{horn1965rationale} (orange line), and Kaiser's criterion threshold (dotted line at $\text{Eigenvalues} = 1$). Here, the scree plot elbow heuristic suggests $Q=4$, whereas parallel analysis indicates $Q=3$. Kaiser's rule yields an ambiguous result: although the fourth eigenvalue falls slightly below unity at approximately $0.92$, discarding it risks omitting a nontrivial proportion of total variance at such a low dimensionality. The Interpretability Index resolves this ambiguity through Figure \ref{fig:wine_chooseQ} (right), which tracks marginal changes in $I$, $\operatorname{ECV}$ and $\overline{\mathrm{GS}}_w$. Going from $Q=3$ to $Q=4$ yields negligible gains in factor simple structure ($\overline{\mathrm{GS}}_w$) and only a marginal increase in overall interpretability ($I$). A fourth factor therefore introduces model complexity without meaningful gains in conceptual clarity, establishing $Q=3$ as the preferred choice.

% =========================================================================
\section{Conclusions}
\label{sec:conclusions}
% =========================================================================

This paper introduced the Interpretability Index $I$ to enhance the evaluation of Exploratory Factor Analysis (EFA) solutions. The proposed metric captures the trade-off between Explained Common Variance ($\operatorname{ECV}$) and a novel aggregated sparseness measure ($\overline{\mathrm{GS}}_w$). This sparseness metric generalizes Hoyer's formulation to variable norm orders $(p_1, p_2)$, better adapting to the analytical notion of a ``negligible loading''. Each factor's contribution to overall sparseness is weighted by its share of information, constructed to yield an exact partition of the total $\operatorname{ECV}$. An asymptotic analysis of $(p_1, p_2)$, combined with a statistical interpretation of candidate norms, establishes $(p_1, p_2) = (2, 4)$ as the canonical parameter choice.

To support the primary index in cases of nearly tied interpretability scores, a complementary row-compliance procedure was developed to identify the best solution among candidate optima. The entire theoretical framework is implemented through comprehensive, easy-to-use computational routines that enable users to immediately apply the methodology and evaluate model improvements. Real-data applications demonstrate the capabilities of this decision framework in identifying optimal factor rotations and resolving ambiguities in selecting the number of factors $Q$ alongside traditional tools.

Finally, this methodology was developed specifically with EFA because its rotational indeterminacy provides the most direct setting for developing and evaluating the proposed framework. Future research will extend this framework to broader dimensionality reduction techniques, where the structural constraints inherent to methods like Principal Component Analysis will unlock new theoretical properties and applications.

\bibliography{references}

\end{document}